\documentstyle[eqsecnum,aps]{revtex}         

\begin{document}


\title{Dimensional Transmutation and
 Dimensional Regularization
\\
 in Quantum Mechanics
\\
{\small\bf II. Rotational Invariance}}

\author{Horacio E. Camblong}
\address{Department of Physics, University of San Francisco, San
Francisco, California 94117-1080
\\
\medskip
{\rm and}                            
}

\author{Luis N. Epele, Huner Fanchiotti,
 and Carlos A. Garc\'{\i}a Canal.}

\address{Laboratorio de F\'{\i}sica Te\'{o}rica,
 Departmento de F\'{\i}sica,
Universidad Nacional de La Plata,
\\ C.C. 67, 1900 La Plata, Argentina
\\
\bigskip
{\sc Dedicated to C.G. Bollini and J.J. Giambiagi} }

\def\diagram#1{{\normallineskip=8pt
       \normalbaselineskip=0pt \matrix{#1}}}

\def\diagramrightarrow#1#2{\smash{\mathop{\hbox to 
.8in{\rightarrowfill}}
        \limits^{\scriptstyle #1}_{\scriptstyle #2}}}

\def\diagramleftarrow#1#2{\smash{\mathop{\hbox to .8in{\leftarrowfill}}
        \limits^{\scriptstyle #1}_{\scriptstyle #2}}}

\def\diagramdownarrow#1#2{\llap{$\scriptstyle #1$}\left\downarrow
    \vcenter to .6in{}\right.\rlap{$\scriptstyle #2$}}

\def\diagramuparrow#1#2{\llap{$\scriptstyle #1$}\left\uparrow
    \vcenter to .6in{}\right.\rlap{$\scriptstyle #2$}}

\maketitle

\begin{quotation}
{\small
A thorough analysis is presented of the class of central fields of
force that exhibit: (i) dimensional transmutation and (ii)
rotational invariance. Using dimensional regularization, the
two-dimensional delta-function potential and the $D$-dimensional
inverse square potential are studied. In particular, the following
features are analyzed: the existence of a critical coupling, the
boundary condition at the origin, the relationship between the
bound-state and scattering sectors, and the similarities displayed
by both potentials. It is found that, for rotationally symmetric
scale-invariant potentials, there is a strong-coupling regime, for
which quantum-mechanical breaking of symmetry takes place, with
the appearance of a unique bound state as well as of a logarithmic
energy dependence of the scattering with respect to the energy.
}
\end{quotation}



\section{INTRODUCTION}
\label{sec:intro}

Dimensional transmutation~\cite{col:73} has become a standard
concept for the analysis of quantum field theories devoid of
intrinsic dimensional parameters. In the first paper in this
series~\cite{cam:00a}, we have shown that dimensional
transmutation is a more general phenomenon related to dimensional
analysis and renormalization, and we provided a general theory for
its description in nonrelativistic quantum mechanics. The main
purpose of this paper is to apply the general theory developed in
Ref.~\cite{cam:00a} to an exhaustive analysis of examples that
exhibit rotational invariance. Parenthetically, as shown in
Ref.~\cite{cam:00a}, scale-invariant potentials---those that
possess scale symmetry at the classical level and do not exhibit
any explicit dimensional scale---are exactly described by
homogeneous functions of degree $-2$; this class includes contact
potentials as well as ordinary potentials of the form $1/r^{2}$,
with a possible angular dependence. Thus, within the subset of
such potentials with rotational invariance, the following two
problems stand out: the two-dimensional delta-function
potential~\cite{tho:79,jac:91,alb:88,cam:00_delta} and the inverse
square
potential~\cite{mot:49,cas:50,mor:53,lan:77,fra:71,par:73,sch:76,sim:70,all:72,gup:93}.
In addition to solving these potentials, we will expand the
general framework presented in~\cite{cam:00a} in order to
encompass further understanding of the scattering
sector---including a partial-wave analysis---as well as a
reexamination of the peculiar boundary conditions satisfied by
these potentials at the origin.

The plan of this paper is as follows.
In Section~\ref{sec:delta}
we discuss the two-dimensional delta-function potential
and compare the results with those earlier derived in 
Ref.~\cite{cam:00a}.
In Section~\ref{sec:ISP_intro}
we summarize the known properties of the inverse
square potential (but generalizing them to $D$ dimensions)
and examine the issue of the boundary condition at the
origin; the conclusion of that section is that
renormalization is needed in the strong-coupling regime.
In Section~\ref{sec:ISP_renormalization} we provide
the required dimensional renormalization of the
inverse square potential and find a complete solution to the problem
by means of a duality transformation.
Section~\ref{sec:conclusions} summarizes our main conclusions and 
outlines a
strategy for answering additional questions.
The appendices explicitly use the
concept of rotational invariance and
deal with the $D$-dimensional central-force problem,
the $D$-dimensional partial-wave expansion,
and the duality transformation.

We conclude this introduction with
some comments about notation and background,
with which we assume that the reader has some familiarity
from Ref.~\cite{cam:00a}.
In particular, in this paper, we will solve the
Schr\"{o}dinger equation
associated with
\begin{equation}
V({\bf r})= - \lambda  \,  W({\bf r})
\;  ,
\label{eq:W}
\end{equation}
for the two-dimensional delta-function potential
\begin{equation}
W({\bf r})=
\delta^{(2)}({\bf r})
\;
\label{eq:2D_delta}
\end{equation}
(Section~\ref{sec:delta}),
and for the inverse square potential
\begin{equation}
W({\bf r})=
\frac{1}{r^{2}}
\;
\label{eq:ISP}
\end{equation}
(Sections~\ref{sec:ISP_intro} and \ref{sec:ISP_renormalization}).
The regularized solutions
will be obtained with dimensional continuation from
$D_{0}$ to $D$ dimensions and
will depend on the parameter $\epsilon
=D_{0}-D$; from these solutions,
we will identify the energy generating function $\Xi (\epsilon) $,
by comparison with the master eigenvalue equation
\begin{equation}
 \lambda \, \mu^{\epsilon} \,
\Xi (\epsilon)
\, | E  (\epsilon) |^{-\epsilon/2} = 1
 \; ,
\label{eq:eigenvalues_general2}
\end{equation}
where $\lambda$ is the dimensionless coupling constant,
$E$ the energy,
and $\mu$ the renormalization scale
that arises from the bare coupling $\lambda_{B}= \mu^{\epsilon} 
\lambda$.
The ensuing analysis of existence of bound states and related
concepts will often refer to properties
of the energy generating function $\Xi (\epsilon) $
and of the critical couplings
\begin{equation}
\lambda^{(\ast)}_{n} = \left[
\lim_{\epsilon\rightarrow 0}
\Xi_{n}(\epsilon)
\right]^{-1}
\;  ,
\label{eq:lambda_star}
\end{equation}
as  discussed in Ref.~\cite{cam:00a}, even though the derived
formulas can mostly be analyzed on their own right and, for the
most part, this paper is essentially self-contained.

\section{TWO-DIMENSIONAL DELTA-FUNCTION POTENTIAL}
\label{sec:delta}

The delta-function potential belongs to
the class of pseudopotentials that
arises in the low-energy limit of
effective quantum field theory~\cite{cam:00b}.
As in Ref.~\cite{cam:00a},
in this section we will use dimensional
regularization and will focus only on those features
that are associated with the dimensional transmutation
of the two-dimensional representative of this class,
Eqs.~(\ref{eq:W}) and (\ref{eq:2D_delta}).

The two-dimensional delta-function potential has been extensively
studied in the literature using a variety of regularization
techniques: (i) momentum-cutoff
regularization~\cite{tho:79,jac:91,hen:97}, (ii) real-space
short-distance regularization (using a circular-well
potential)~\cite{hol:93,mea:91,gos:91,man:94}, (iii) Pauli-Villars
regularization~\cite{mit:98}, (iv) dimensional
regularization~\cite{cam:00a,mit:98}, and (v) method of
self-adjoint extensions~\cite{jac:91,alb:88}. As expected, the
final renormalized results are independent of the regularization
scheme, despite the dissimilar appearance of the regularized
expressions. Even though our treatment will be solely based on
dimensional regularization, the establishment of a peculiar
boundary condition at the origin shows a natural connection with
method (v), where the singular nature of the potential at the
origin is replaced by a boundary condition in order to preserve
the self-adjoint character of the Hamiltonian.

 The dimensionally regularized
Schr\"{o}dinger equation for the potential~(\ref{eq:2D_delta}) can
be solved in a number of different ways. In Ref.~\cite{cam:00a} we
solved it in momentum space---a procedure that works due to the
zero-range nature of this potential~\cite{tho:79}. The same
feature permits an effective solution in hyperspherical
coordinates as if it were a central potential (according to the
framework developed in Appendix~\ref{sec:central}). In fact,
 the similarities between the two-dimensional
delta-function and inverse square potentials will become more 
transparent,
by the use of hyperspherical coordinates for both.

According to the dimensional-continuation prescription
of Ref.~\cite{cam:00a}, we Fourier transform
Eq.~(\ref{eq:2D_delta}) in $D_{0}=2$ dimensions,
then perform a dimensional jump
to $D$ dimensions in Fourier space, and finally return
to position space in $D$ dimensions, with the obvious result
\begin{equation}
[- \nabla_{ {\bf r}, D }^2   -
    \lambda \, \mu^{\epsilon} \,
\delta^{({D})}
 ( {\bf r}  )  ]
\,  \Psi (  {\bf r}  )
= E
 \, \Psi (  {\bf r}  )
\;  .
\label{eq:Schr2nu1delta}
\end{equation}

From the generalized angular-momentum analysis for central
potentials of Appendix~\ref{sec:central}, it follows that the
radial part of Eq.~(\ref{eq:Schr2nu1delta}) in hyperspherical
coordinates reduces to the homogeneous form
\begin{equation}
\left[ \frac{d^{2}}{dr^{2}}
+ E -  \frac{(l + \nu)^{2}
 - 1/4 }{r^{2}}
\right]  u_{l}(r) = 0
\;  ,
\label{eq:radialSchr_delta}
\end{equation}
for any $r \neq 0$.
In Eq.~(\ref{eq:radialSchr_delta}), as well as
in our subsequent analysis, both for the two-dimensional
delta-function and inverse square potentials,
the number $D$ of dimensions will usually
appear in terms of the  variable
\begin{equation}
\nu= D/2 -1
\; ,
\label{eq:nu_def}
\end{equation}
which will thereby simplify the form of most formulas.

Equation~(\ref{eq:radialSchr_delta})
is indistinguishable from a free particle
except for the stringent boundary condition
enforced by the delta-function singularity at the origin.
We now move on to analyze this boundary condition.

\subsection{Boundary Condition
at the Origin for the Two-Dimensional Delta-Function Potential}
\label{sec:delta_BC}

Simple inspection of Eqs.~(\ref{eq:Schr2nu1delta})
and (\ref{eq:radialSchr_delta}) shows that
the delta-function singularity represents the only
difference between them.
Even though the correct equation is~(\ref{eq:Schr2nu1delta}),
one can still safely use~(\ref{eq:radialSchr_delta}),
provided that the singularity be replaced
by an appropriate boundary condition at the origin.
This can be established
by means of the small-argument behavior of the
wave function
(neglecting the energy term as $r \rightarrow 0$)
\begin{equation}
-\nabla_{ {\bf r},  D }^2
\Psi( {\bf r})
\stackrel{\scriptscriptstyle
(r \rightarrow 0)}{\sim}
 \lambda   \, \mu^{\epsilon}
 \,
\Psi ({\bf 0})
\,
\delta^{(D)} ({\bf r})
\;  ,
\label{eq:limiting_delta_singularity}
\end{equation}
which can be evaluated by comparison with the identity
\begin{equation}
-\nabla_{ {\bf r},  D }^2
\left[
\frac{ r^{-(D-2)}}{(D-2) \Omega_{D}}
\right]
=
\delta^{(D)} ({\bf r})
\;  .
\label{eq:Laplace_GF_delta}
\end{equation}
Equations~(\ref{eq:limiting_delta_singularity})
and (\ref{eq:Laplace_GF_delta})
 imply that
the general form of
$\Psi ({\bf r})$ near the origin is
\begin{equation}
\Psi ({\bf r})
\stackrel{\scriptscriptstyle
(r \rightarrow 0)}{\sim}
\Psi ({\bf 0})
\frac{\lambda \, \mu^{\epsilon} }{ (D-2) \Omega_{D} }  r^{-(D-2)}
+ \Psi_{h} ({\bf r})
\; ,
\label{eq:delta_BC1}
\end{equation}
where $\Psi_{h} ({\bf r})$
is the general solution of the homogeneous equation.
Moreover,
Eq.~(\ref{eq:delta_BC1}) can be made more explicit
by means of
\begin{equation}
\Psi_{h} ({\bf r})
 =  \sum_{l=0}^{\infty}
\left[ C^{(+)}_{l} r^{l} + C^{(-)}_{l}
r^{-(l+2 \nu)} \right] Y_{L}(\Omega^{(D)})
\;
\end{equation}
and
\begin{equation}
-\nabla_{ {\bf r},  D }^2
\left[  r^{-(l+2\nu)} Y_{L}(\Omega^{(D)})
\right]
\propto
\nabla^{l}
\delta^{(D)} ({\bf r})
\;  ,
\label{eq:D-dimensional_multipoles}
\end{equation}
with $\nabla^{l} $ being a certain linear combination of
$l$th order derivatives; then,
it follows that
$C^{(-)}_{l}  =0$ for all $l$,
for otherwise $C^{(-)}_{l}$ would modify the
inhomogeneous part of Eq.~(\ref{eq:delta_BC1}).
Then, resolving the wave function into individual angular
momentum components,
\begin{equation}
\Psi ({\bf r})
=  \sum_{l=0}^{\infty}
R_{l} (r) \,
Y_{L}(\Omega^{(D)})
\;  ,
\end{equation}
we have the following set of boundary conditions.
First, for $l=0$,
the value of the constant $ C^{(+)}_{0} $
corresponds to $\Psi ({\bf 0})$, as required by
 self-consistency of the value of the
wave function at the origin; then,
\begin{equation}
R_{0} ( r)
 \stackrel{\scriptscriptstyle
(r \rightarrow 0)}{\sim}
R_{0} ( 0)
\,
\left[
\frac{\lambda \, \mu^{-2 \nu}
\, \Gamma (\nu)}{4 \pi^{\nu +1}} \,
 r^{-2 \nu} + 1
\right]
\; .
\label{eq:delta_BC2}
\end{equation}
In addition,
for $l \neq 0$,
\begin{equation}
R_{l} ( r)
 \stackrel{\scriptscriptstyle
(r \rightarrow 0)}{\sim}
 C^{(+)}_{l}
\,
r^{l}
\; .
\label{eq:delta_BC2_l_neq_0}
\end{equation}

Equations~(\ref{eq:delta_BC2})
and (\ref{eq:delta_BC2_l_neq_0})
are the required boundary conditions at the origin.
In particular,
Eq.~(\ref{eq:delta_BC2})
implies that
 $D<2$ is the condition for regularity,
while the case $D=2$ is ``critical.''

\subsection{Bound-State Sector for a
Two-Dimensional Delta-Function Potential}
\label{sec:delta_bound_states}

The coupling $\lambda$ in Eq.~(\ref{eq:2D_delta})
defines an ``attractive'' zero-range potential
for $\lambda >0$ and a ``repulsive'' one for
 $\lambda <0$.
By the form of the potential,
this physical argument implies that all states with $E>0$ are of the
scattering type, while states with $E<0$ can only
be bound and are impossible for
$\lambda <0$.
In other words,
there exists a critical coupling
\begin{equation}
\lambda^{(\ast)}=0
\; ,
\label{eq:delta_critical_coupling}
\end{equation}
which separates the theory into two regimes.
Therefore, for a delta-function potential,
the strong-coupling regime ($\lambda>\lambda^{(\ast)}=0$)
coincides with the set of attractive potentials,
while the weak-coupling regime ($\lambda<\lambda^{(\ast)}=0$)
amounts to repulsive potentials.

Let us now consider the bound-state sector, with energy
 $E=-\kappa^{2} <0$,
for an attractive two-dimensional delta-function potential.
The corresponding solution of
Eq.~(\ref{eq:radialSchr_delta}) is
\begin{equation}
\frac{u_{l}(r)}{\sqrt{r}}
=
\mbox{\boldmath\large  $\left\{  \right.$ } \! \! \!
I_{l+\nu}(\kappa r)
\mbox{\boldmath\large  $,$ } \!   \!
K_{l+\nu} (\kappa r)
\mbox{\boldmath\large  $\left.  \right\}$ } \! \!
\;  ,
\label{eq:delta_weak_BS}
\end{equation}
where the symbol
$\mbox{\boldmath\large  $\left\{  \right.$ } \! \! \!
\mbox{\boldmath\large  $,$ } \! \! \!
\mbox{\boldmath\large  $\left.  \right\}$ } \! \!
$
stands for linear combination,
and
$I_{p}(z)$ and $K_{p}(z)$ are
the modified Bessel functions of the first and second kinds,
respectively~\cite{abr:72}.

Of course, the boundary conditions restrict the
selection in Eq.~(\ref{eq:delta_weak_BS}).
First, the boundary condition
at infinity  leads to the rejection of
the modified Bessel function of the first kind,
so that (from Eq.~(\ref{eq:effective_radial_wf}))
\begin{equation}
R_{l}(r)
=
A_{l}
\,
r^{-\nu}
\,
K_{l+\nu} (\kappa r)
\;  .
\label{eq:delta_BS_wf}
\end{equation}

Next, the boundary condition at the origin,
Eqs.~(\ref{eq:delta_BC2}) and (\ref{eq:delta_BC2_l_neq_0}), can be 
enforced
at the level of  Eq.~(\ref{eq:delta_BS_wf})
by considering the small-argument behavior of
the modified Bessel function of the second kind~\cite{abr:72,cam:00c},
\begin{equation}
  K_{p} (z)
  \stackrel{(z \rightarrow 0)}{\sim}
 \frac{1}{2}
\,
\left[
\Gamma (p)
\left( \frac{z}{2} \right)^{-p}
 +
\Gamma (-p)
\left( \frac{z}{2} \right)^{p}
\right]
\,
\left[ 1 + O(z^{2}) \right]
\;  .
\label{mod_Bessel_small_arg}
\end{equation}
For $l \neq 0$,
Eq.~(\ref{eq:delta_BS_wf}) gives a singular term
proportional to $r^{-p}$, with $p=l+\nu$, according to
Eq.~(\ref{mod_Bessel_small_arg}).
Therefore,
the boundary condition
can {\em only\/} be satisfied for
\begin{equation}
l=0
\;  ,
\label{eq:delta_s_states}
\end{equation}
so that the delta-function potential, being of zero range,
can only sustain bound states in the absence of a centrifugal barrier
($s$ states).
Then, for the radial wave function with $l=0$,
\begin{eqnarray}
R_{0} (r)
& = & A_{0}
\,
r^{-\nu}
\, K_{\nu} (\kappa r)
\label{eq:delta_BS_wf_l=0}
\\
&
\stackrel{\scriptscriptstyle (r \rightarrow 0) }{\sim}
&
\frac{A_{0}}{2}
\left( \frac{\kappa}{2} \right)^{\nu}
\left[
\Gamma (-\nu) +
\Gamma (\nu)    \left( \frac{\kappa r}{2} \right)^{-2 \nu}
\right]
\left[
1 + O(r^{2}) \right]
\; .
\label{eq:delta_BC3}
\end{eqnarray}
In conclusion,
compatibility of Eqs.~(\ref{eq:delta_BC2})
and (\ref{eq:delta_BC3})
requires that the following two conditions be simultaneously met:
(i) that
\begin{equation}
A_{0} = 2
R_{0} (0)
\,
\left( \frac{2}{\kappa} \right)^{-\nu}
\,
\frac{1}{ \Gamma (-\nu)}
\; ,
\label{eq:delta_BC4}
\end{equation}
which just enforces the condition of finiteness at the origin
and requires $D<2$ for regularity;
and (ii)
that the bound-state energies $E$
satisfy the eigenvalue condition
\begin{equation}
\frac{\lambda \, \mu^{-2 \nu} }{4\pi}
\left(
\frac{|E|}{4\pi}
\right)^{\nu}
\Gamma
\left( -\nu \right) = 1
\; .
\label{eq:delta_eigen3}
\end{equation}
Equations~(\ref{eq:delta_BC3}),
(\ref{eq:delta_BC4}),
and (\ref{eq:delta_eigen3})
are in complete agreement with the
results obtained in Ref.~\cite{cam:00a}, to which we refer
the reader for additional comments.

As there is no
other quantum number in this problem, and $l=0$ is required,
we see that Eq.~(\ref{eq:delta_BS_wf_l=0}) represents the ground
state of the regularized system.
Parenthetically, the proportionality
constant $A_{0}$ can be found by normalization
by means of the identity~\cite{gra:80}
\begin{equation}
\int_{0}^{\infty} x \left[
K_{\nu}(x) \right]^{2}
d x
=
\frac{1}{2} \, \Gamma(1+\nu) \, \Gamma(1-\nu)
\; ,
\end{equation}
so that
\begin{equation}
\Psi ({\bf r})
=
\frac{\kappa}{ \pi^{(\nu+1)/2} [\Gamma (1-\nu) ]^{1/2} }
 \,
\frac{
K_{\nu} (\kappa r)
}
{r^{\nu}  }
\;  .
\label{eq:delta_wf_normalized}
\end{equation}
Equations~(\ref{eq:delta_eigen3}) and
(\ref{eq:delta_wf_normalized}) indeed reduce to the known
expressions for $D=1$~\cite{rob:97}. Here, no attempt is made to
draw any conclusions about the cases $D>2$, as they require a
separate regularization procedure.

As discussed in Ref.~\cite{cam:00a},
Eq.~(\ref{eq:delta_eigen3})
is equivalent to the master eigenvalue
equation~(\ref{eq:eigenvalues_general2}),
with an energy generating function
\begin{equation}
\Xi (\epsilon) =
\frac{1}{4\pi} \left(4\pi\right)^{\epsilon/2 }
\Gamma
\left(\frac{\epsilon}{2}
\right)
\; ,
\label{eq:delta_master_eigenvalue_function}
\end{equation}
which, not having any discrete labels,
produces just a ground state
\begin{equation}
E_{_{\rm (gs)}}
=
- \mu^{2}
e^{ g^{(0)}  }
\leadsto
 - \mu^2
\;  ,
\label{delta_gs}
\end{equation}
but  no excited states.
In Eq.~(\ref{delta_gs}), the
symbol $\leadsto$ refers to the freedom to make the choice
$g^{(0)} =0 $, which means
no loss of generality, as both
$g^{(0)}  $ and $\mu$ are totally arbitrary~\cite{cam:00a}.
Moreover, the critical coupling is
$
\lambda^{(\ast)}=0
$,
as anticipated earlier by the argument leading to
Eq.~(\ref{eq:delta_critical_coupling});
more precisely, the coupling constant
is asymptotically critical according to the scheme
\begin{equation}
\lambda(\epsilon) = 2 \pi \epsilon
\left\{ 1 + \frac{\epsilon}{2}  \,
\left[
g^{(0)}
- \left(  \ln 4\pi - \gamma \right)
\right]
\right\}
\; ,
\label{eq:delta_renormalized_coupling}
\end{equation}
where $\gamma$ is the Euler-Mascheroni constant.

Finally, the ground state wave function
of the renormalized two-dimensional delta-function potential
can be obtained in the limit $\nu \rightarrow 0$ of
Eq.~(\ref{eq:delta_wf_normalized}),
and using Eq.~(\ref{delta_gs}),
with the result
\begin{equation}
\Psi_{_{\rm (gs)}}
 ({\bf r})
=
\frac{\mu}{ \sqrt{\pi} } \, K_{0} (\mu r)
 \; ,
\label{eq:delta_wf_normalized_renormalized}
\end{equation}
which was derived by using a renormalized path integral in
Ref.~\cite{hen:97}.

\subsection{Scattering Sector for a Two-Dimensional
Delta-Function Potential}
\label{sec:delta_scattering}

Scattering under the action of a two-dimensional
delta-function potential
will also be analyzed
by an explicit partial-wave resolution  in
hyperspherical coordinates according to the theory
of Appendix~\ref{sec:D-dim_partial_wave_analysis}.
The first step is to solve
Eq.~(\ref{eq:radialSchr_delta}), which for the scattering sector,
$E=k^{2}>0$, admits the solution
\begin{equation}
u_{l} (r)
= \sqrt{r}
\left[
A_{l}^{(+)} H_{l+\nu}^{(1)} (kr)
+
A_{l}^{(-)} H_{l+\nu}^{(2)} (kr)
\right]
\;  ,
\label{eq:delta_scatt_exact_sol}
\end{equation}
where the $H_{p}^{(1,2)} (z)$ are Hankel functions,
whose behavior near the origin implies the
asymptotic dependence
\begin{equation}
u_{l} (r)
\stackrel{\scriptscriptstyle
(r \rightarrow 0)}{\sim}
\frac{ \sqrt{r}}{\pi i }
\left\{
\left[
A_{l}^{(+)} e^{-i \pi (l+\nu)}
-
A_{l}^{(-)} e^{i \pi (l+\nu)}
\right]
\Gamma (-\nu)  \left( \frac{kr}{2} \right)^{l+\nu}
+
\left[
A_{l}^{(+)}
-
A_{l}^{(-)}
\right]
\Gamma (\nu)  \left( \frac{kr}{2} \right)^{-(l+\nu)}
\right\}
\;  .
\label{eq:delta_scatt_near_origin}
\end{equation}
Equations~(\ref{eq:delta_scatt_near_origin}) and
(\ref{eq:effective_radial_wf})
yield
\begin{eqnarray}
R_{l} (r)
&
\stackrel{\scriptscriptstyle
(r \rightarrow 0)}{\sim}
&
 \frac{A^{(-)}}{\pi i}
\left[
S^{(D)}_{l}(k)
e^{-i \pi (l+\nu)}
-
e^{i \pi (l+\nu)}
\right]
\Gamma (-\nu)  \left( \frac{kr}{2} \right)^{l}
\nonumber \\
&  \times &
 \left\{
1 +
\frac{
\left[ S^{(D)}_{l}(k) -1 \right]
\,  \Gamma (\nu) \, \left( 2/k \right)^{2\nu}    }
{
\left[
S^{(D)}_{l}(k)
 \, e^{-i \pi (l+\nu)}
-
e^{i \pi (l+\nu)}
\right]
\,
\Gamma (-\nu)
}
\;
r^{-\left( 2 l+2 \nu \right) }
\right\}
\;  ,
\label{eq:delta_BC_scatt}
\end{eqnarray}
where the scattering matrix elements
$S^{(D)}_{l}(k) $ have been explicitly introduced
by means of Eq.~(\ref{eq:scatt_matrix_from_coeff}).

Equation~(\ref{eq:delta_BC_scatt})
should be compared again with the
boundary condition at the origin.
According to Eq.~(\ref{eq:D-dimensional_multipoles}),
for $l \neq 0$, the boundary condition~(\ref{eq:delta_BC2_l_neq_0})
cannot be satisfied, unless
\begin{equation}
\left.
S^{(D)}_{l}(k)
\right|_{l \neq 0}
=
1
\;  ;
\label{eq:delta_scatt_matrix_l_neq_0}
\end{equation}
in particular, for the phase shifts,
\begin{equation}
\left.
\delta^{(D)}_{l}(k)
\right|_{l \neq 0}
=
0
\;  ,
\end{equation}
with the conclusion that there is no scattering for $l \neq 0$. On
the other hand, for the $s$ wave, Eqs.~(\ref{eq:delta_BC2}) and
(\ref{eq:delta_BC_scatt}) yield a scattering matrix element
\begin{equation}
S^{(D)}_{0}(k)
=
\frac{
1 - \lambda \,  e^{i \pi \nu} \, \Xi (-2 \nu) \,
     \left(E/\mu^{2}\right)^{\nu}
}
{
1 -  \lambda \, e^{-i \pi \nu}\,  \Xi (-2 \nu) \,
\left(E/\mu^{2}\right)^{\nu}
}
\;  ,
\label{eq:delta_scatt_matrix}
\end{equation}
where
\begin{equation}
\Xi (-2  \nu)
=
\frac{\Gamma(-\nu)}{(4 \pi )^{\nu + 1}}
\;
\label{eq:delta_master_eigenvalue_function2}
\end{equation}
is identical to
the energy generating function,
Eq.~(\ref{eq:delta_master_eigenvalue_function}).
Equation~(\ref{eq:delta_scatt_matrix}) provides the phase shift
$\delta^{(D)}_{0}(k) $ through Eq.~(\ref{eq:scatt_matrix_def}), whence
\begin{equation}
\tan
\delta^{(D)}_{0}(k)
=
\tan \pi \nu \,
\left[
1-
\frac{\sec \pi \nu }{ \lambda \,
\left( k/\mu \right)^{2 \nu} \,  \Xi (-2 \nu) }
\right]^{-1}
\;  .
\end{equation}

We are now ready to derive the expressions for the
two-dimensional delta-function potential.
First, for $\lambda<0$ (repulsive potential),
$\lambda$ remains unrenormalized and the limit
$\epsilon = - 2 \nu \rightarrow 0$
yields
\begin{equation}
\left.
S^{(D)}_{0}(k)
\right|_{\lambda<0}
=
1
\;  ,
\label{eq:delta_scatt_matrix_repulsive}
\end{equation}
so that there is no scattering whatsoever for repulsive
two-dimensional delta-function potentials.
Instead, for $\lambda>0$ (attractive potential), in the limit
$\epsilon = - 2 \nu \rightarrow 0$
these expressions should be renormalized by means  of
Eq.~(\ref{eq:delta_renormalized_coupling}),
whence
\begin{equation}
\left.
S^{(2)}_{0}(k)
\right|_{\lambda>0}
=
\frac{
\ln \left( E/\mu^{2} \right) - g^{(0)} + i \pi
} { \ln \left( E/\mu^{2} \right)  - g^{(0)} -
i \pi }
 \; .
 \label{eq:delta_scatt_matrix_repulsive2}
\end{equation}
Equation~(\ref{eq:delta_scatt_matrix_repulsive2}) can be further
simplified with Eq.~(\ref{delta_gs}), leading to
\begin{equation}
S^{(2)}_{0}(k)
=
\frac{
\ln \left(k^{2}/|E_{_{\rm (gs)}}| \right) + i \pi
}
{
\ln \left(k^{2}/|E_{_{\rm (gs)}}| \right) - i \pi
}
\;  ,
\label{eq:2D_delta_scatt_matrix}
\end{equation}
and~\cite{mea:91,gos:91,ben:99,cav:99}
\begin{equation}
\tan
\delta^{(2)}_{0}(k)
=
\frac{\pi}{
\ln \left( k^{2}/|E_{_{\rm (gs)}}| \right) }
\;  ,
\label{eq:2D_delta_phase_shift}
\end{equation}
where, from this point on, it will
be understood that scattering is nontrivial only
for $\lambda>0$.
In particular,
from Eqs.~(\ref{eq:delta_scatt_matrix_l_neq_0}),
(\ref{eq:2D_delta_scatt_matrix}),
(\ref{eq:N_nu}), (\ref{eq:scatt_amplitude}),
and (\ref{eq:partial_wave_amplitude}),
the scattering amplitude
becomes
\begin{eqnarray}
f^{(2)}_{k}
( \Omega^{(2)}   )
& =  &
\sqrt{\frac{2}{\pi k}}
\,
\frac{  S^{(2)}_{0}(k)
- 1}{2 i}
\nonumber \\
& = &
\sqrt{\frac{2\pi}{ k}} \,
\left[ \ln \left( \frac{k^{2}}{E_{_{\rm (gs)}}} \right)
-i \pi
\right]^{-1}
\;  .
\label{eq:delta_scattering_amplitude}
\end{eqnarray}
Equation~(\ref{eq:delta_scattering_amplitude}) is identical to the
corresponding expression derived in Ref.~\cite{cam:00a} and
coincides with the corresponding expressions derived in
Refs.~\cite{hol:93,mea:91,man:94,mit:98,ben:99,cav:99}.

A number of consequences follow from
Eqs.~(\ref{eq:2D_delta_scatt_matrix}),
(\ref{eq:2D_delta_phase_shift}),
and (\ref{eq:delta_scattering_amplitude}):
\begin{enumerate}
\item
 The unique pole of the scattering
matrix~(\ref{eq:2D_delta_scatt_matrix})
corresponds to the unique bound state.
\item
The phase shifts can be alternatively renormalized using a
floating scale $\mu$, in the form
\begin{equation}
\frac{1}{\tan \delta^{(2)}_{0}(k)
 }
=
\frac{1}{\tan \delta^{(2)}_{0}(\mu)  }
+ \frac{1}{\pi}
\ln  \left( \frac{k}{\mu} \right)^{2}
\; .
\label{delta_scatt_floating_renormalization}
\end{equation}
\item
Equation~(\ref{eq:delta_scattering_amplitude})
 yields  a differential scattering cross section
\begin{equation}
\frac{d \sigma^{(2)}
(k, \Omega^{(2)} )}{d
\Omega_{2}}
=
\frac{2\pi}{ k} \,
\left\{
\left[ \ln \left( \frac{k^{2}}{E_{_{\rm (gs)}}}
\right) \right]^{2}
+ \pi^{2}
\right\}^{-1}
\;  .
\label{eq:delta_diff_scatt}
\end{equation}
\item
The total scattering cross section becomes
\begin{equation}
\sigma_{2}(k)
=
\frac{4 \pi^{2}}{k}
\,
\left\{
\left[
\ln \left( E/|E_{_{\rm (gs)}}| \right)
 \right]^{2}
+ \pi^{2}
\right\}^{-1}
\;  .
\label{eq:delta_total_scatt}
\end{equation}
\item
All the relevant quantities are logarithmic with respect to
the energy and agree with the predictions of
generalized dimensional analysis~\cite{cam:00a}.
\item
Equations~(\ref{eq:2D_delta_scatt_matrix}),
(\ref{eq:2D_delta_phase_shift}),
(\ref{eq:delta_scattering_amplitude}),
(\ref{eq:delta_diff_scatt}), and (\ref{eq:delta_total_scatt})
relate the bound-state and scattering sectors of the theory and
confirm that the two-dimensional
delta-function potential is renormalizable.
\end{enumerate}

\section{INVERSE SQUARE POTENTIAL: INTRODUCTION}
\label{sec:ISP_intro}

It is well known that
the inverse square potential, $V({\bf r})\propto r^{-2}$,
displays a number of unusual
features in its quantum-mechanical version.
In a sense, it represents the boundary between regular
and singular power-law potentials (Appendix~\ref{sec:powerlaw}).
As we will see in this section,
its marginally singular nature can be traced back to its interplay with
the centrifugal barrier and produces
two regimes separated by a critical coupling.
These features have become part of the standard background on
central potentials in quantum mechanics~\cite{mot:49,mor:53,lan:77}.
The difficulties encountered in the strong-coupling regime
correspond to the classical picture
of the ``fall of the particle to the center''~\cite{lan:77}.
It should be pointed out that
this problem is relevant in polymer physics~\cite{mar:91},
as well as in molecular
physics, where it appears in a modified form as a dipole
potential~\cite{lev:67,des:94}.

In this section, our goal is to
review the origin of
this singular behavior
and pave the way for the regularization and renormalization
of the theory, which we will implement in
Section~\ref{sec:ISP_renormalization}.

\subsection{Exact Solution for the Unregularized Inverse Square 
Potential}
\label{sec:ISP_NR}

Stationary eigenstates
of energy and orbital angular momentum
of the unregularized inverse square potential
are of the factorized form~(\ref{eq:separation_Psi}),
with an effective radial function~(\ref{eq:effective_radial_wf})
that satisfies the $D_{0}$-dimensional
radial Schr\"{o}dinger equation
(Eqs.~(\ref{eq:central_eff_Schr})--(\ref{eq:cf_coupling})),
\begin{equation}
\left[ \frac{d^{2}}{dr^{2}} + E -
\frac{ \left( l + \nu_{0} \right)^{2}  - \lambda
- 1/4}{r^{2}}
\right]  u_{l}(r) = 0
\;  ,
\label{eq:ISP_radial}
\end{equation}
where  $\nu_{0}=D_{0}/2-1$
and $\lambda>0$ corresponds to an attractive potential.
Then solutions to Eq.~(\ref{eq:ISP_radial}) are of the form
\begin{equation}
  u_{l}(r)
=
\sqrt{r}
\,
Z_{s_{l}}
\left(\sqrt{E} \, r
\right)
\; ,
\label{eq:ISP_solutions}
\end{equation}
where $Z_{s_{l}} (z) $ represents an appropriate linear combination
of Bessel functions of order
\begin{equation}
s_{l}
=
\sqrt{  \lambda_{l}^{(\ast)}
- \lambda}
\; ,
\label{eq:ISP_order}
\end{equation}
with
\begin{equation}
\lambda_{l}^{(\ast)}
= \left( l +  \nu_{0}  \right)^{2}
\;  .
\label{eq:ISP_critical_coupling}
\end{equation}
It is immediately apparent that
the inverse square potential in Eq.~(\ref{eq:ISP_radial})
has  the same dependence on the radial
variable as the centrifugal potential, so that
its effect is solely to modify the strength of the centrifugal barrier;
correspondingly, the solution~(\ref{eq:ISP_solutions})
is essentially a free-particle wave function where
the replacement
\begin{equation}
(l+\nu_{0}) \rightarrow s_{l}
\;
\label{eq:ISP_order_replacement_from_free_part}
\end{equation}
has been made.
The parameter
$\lambda_{l}^{(\ast)} $ in Eq.~(\ref{eq:ISP_critical_coupling})
plays the role of a critical coupling, i.e.,
the nature of the solutions changes abruptly around the value
$ \lambda =\lambda_{l}^{(\ast)} $, for
any state with angular momentum $l$. Thus,
$\lambda_{l}^{(\ast)} $
acts as the boundary between two
regimes: (i) weak coupling,
characterized by $ \lambda <\lambda_{l}^{(\ast)} $,
for which the order $s_{l} $ is real;
and (ii) strong coupling,
characterized by $ \lambda >\lambda_{l}^{(\ast)} $,
for which the order $s_{l} =i \Theta_{l}$ is imaginary,
with
\begin{equation}
\Theta_{l}
= \sqrt{ \lambda - \lambda_{l}^{(\ast)} }
\;  .
\label{eq:Theta}
\end{equation}
The character of the solutions also depends
on the other relevant parameter
in Eq.~(\ref{eq:ISP_radial}), namely, the energy $E$,
in such a way that: (i)  scattering states are only possible if
$E=k^{2}>0$, with the argument of the Bessel functions
in Eq.~(\ref{eq:ISP_solutions}) being $kr $ (real);
while (ii) bound states  are only possible if
$E=-\kappa^{2} < 0$, with the argument of the Bessel functions
in Eq.~(\ref{eq:ISP_solutions}) being $kr = i \kappa r $
(imaginary). In conclusion,
the solutions~(\ref{eq:ISP_solutions})
fall into the following four families,
according to the nature of the two relevant variables
$s_{l} $ and $E$:
\begin{enumerate}
\item
Bound-state sector
of the weak-coupling regime, with
\begin{equation}
\frac{u_{l}(r)}{\sqrt{r}}
=
\mbox{\boldmath\large  $\left\{  \right.$ } \! \! \!
I_{s_{l}}(\kappa r)
\mbox{\boldmath\large  $,$ } \!   \!
K_{s_{l}} (\kappa r)
\mbox{\boldmath\large  $\left.  \right\}$ } \! \!
\;  ;
\label{eq:ISP_weak_BS}
\end{equation}
\item
Scattering sector
of the weak-coupling regime, with
\begin{equation}
\frac{u_{l}(r)}{\sqrt{r}}
 =
\mbox{\boldmath\large  $\left\{  \right.$ } \! \! \!
H^{(1)}_{s_{l}}(kr)
\mbox{\boldmath\large  $,$ }  \!   \!
H^{(2)}_{s_{l}}(kr)
\mbox{\boldmath\large  $\left.  \right\}$ } \! \!
\;  ;
\label{eq:ISP_weak_scatt}
\end{equation}
\item
Bound-state sector
of the strong-coupling regime, with
\begin{equation}
\frac{u_{l}(r)}{\sqrt{r}}
=
\mbox{\boldmath\large  $\left\{  \right.$ } \! \! \!
I_{i\Theta_{l}} (\kappa r)
\mbox{\boldmath\large  $,$ } \!   \!
K_{i\Theta_{l}} (\kappa r)
\mbox{\boldmath\large  $\left.  \right\}$ } \! \!
\;  ;
\label{eq:ISP_strong_BS}
\end{equation}
\item
Scattering  sector
of the strong-coupling regime, with
\begin{equation}
\frac{u_{l}(r)}{\sqrt{r}}
=
\mbox{\boldmath\large  $\left\{  \right.$ } \! \! \!
H^{(1)}_{i\Theta_{l}}(kr)
\mbox{\boldmath\large  $,$ } \!   \!
H^{(2)}_{i\Theta_{l}}(kr)
\mbox{\boldmath\large  $\left.  \right\}$ } \! \!
\;  ;
\label{eq:ISP_strong_scatt}
\end{equation}
\end{enumerate}
where the symbol
$\mbox{\boldmath\large  $\left\{  \right.$ } \! \! \!
\mbox{\boldmath\large  $,$ } \! \! \!
\mbox{\boldmath\large  $\left.  \right\}$ } \! \!
$
stands again for linear combination.
In Eqs.~(\ref{eq:ISP_weak_BS})--(\ref{eq:ISP_strong_scatt})
we use the standard notation
$H_{s_{l}}^{(1,2)} (z)$ for the Hankel functions,
and $I_{s_{l}}(z)$ and $K_{s_{l}}(z)$
for the modified Bessel functions of the first and second kinds,
respectively~\cite{abr:72}.

Let us first look at the weak-coupling regime.
For bound states,
the requirement that the solution be
finite at infinity implies the rejection of the
component $I_{s_{l}}(\kappa r)$ in Eq.~(\ref{eq:ISP_weak_BS}),
whereas the boundary condition at the origin,
Eq.~(\ref{eq:BC_at_origin}),
leads to the elimination  of the component
$K_{s_{l}}(\kappa r)$.
In other words,
there exist no bound states for a weak coupling.
Physically, this state of affairs is
reminiscent of the nonexistence of bound states for a repulsive
potential;
in fact,
the combination of the potential itself and the centrifugal potential
is  effectively ``repulsive'' when $\lambda<\lambda_{l}^{(\ast)} $.
On the other hand, such a potential can produce nontrivial scattering,
which, according to Eq.~(\ref{eq:ISP_weak_scatt}),
is described by the solution
\begin{eqnarray}
u_{l}(r)
& = &
\sqrt{r}
\,
\left[
\widetilde{A}_{l}^{(+)} H^{(1)}_{s_{l}}(kr)
+
\widetilde{A}_{l}^{(-)}
H^{(2)}_{s_{l}}(kr)
\right]
\nonumber \\
& \stackrel{(r \rightarrow \infty) }{\sim} &
\sqrt{r} \,
\left[
\widetilde{A}_{l}^{(+)}
 H^{(1)}_{l+\nu_{0}}
\mbox{\boldmath\large  $\left(  \right.$ } \! \! \!
k r + \left( l+\nu_{0} -s_{l} \right) \frac{\pi}{2}
\mbox{\boldmath\large  $\left.  \right)$ } \! \!
\right.
\nonumber \\
&  & \;  \;  \;  \;  \;  \;  \;  \;
\left.
+
\widetilde{A}_{l}^{(-)}
 H^{(2)}_{l+\nu_{0}} 
\mbox{\boldmath\large  $\left(  \right.$ } \! \! \!
k r + \left( l+\nu_{0} - s_{l} \right) \frac{\pi}{2}
\mbox{\boldmath\large  $\left.  \right)$ } \! \!
\right]
\;  ,
\label{eq:ISP_weak_scatt_asymptotic_wf}
\end{eqnarray}
where 
$\widetilde{A}_{l}^{(+)}= \widetilde{A}_{l}^{(-)}$
and the asymptotic behavior determines the
energy-independent phase shifts
\begin{equation}
 \delta_{l}^{(D_{0})}
=
 \left[ \left(l+\nu_{0} \right)
- \sqrt{  \left(l+\nu_{0} \right)^{2}  - \lambda}
\right]  \, \frac{\pi}{2}
\; .
\label{ISP_weak_phase_shifts}
\end{equation}
Equation~(\ref{ISP_weak_phase_shifts}) manifestly displays the
scale invariance of the theory, generalizing the well-known
three-dimensional results~\cite{mot:49,jac:72}.
Notice that the scattering matrix $S_{l}^{(D_{0})} =
\exp \left[ 2i \delta_{l}^{(D_{0})} \right] $
has no poles, which is in agreement with
the absence of bound states.
Parenthetically,
our conclusions relied on the choice of the boundary
condition~(\ref{eq:BC_at_origin}) at the origin;
the validity of this procedure for the inverse square potential
will be proved in the next section.

Let us now consider the strong-coupling regime.
For bound states,
the requirement that the solution be
finite at infinity implies again the rejection of the
component $I_{i\Theta_{l}}(\kappa r)$ in Eq.~(\ref{eq:ISP_strong_BS}).
However,
the boundary condition at the origin,
Eq.~(\ref{eq:BC_at_origin}),
can neither eliminate the component
$K_{s_{l}}(\kappa r)$ nor restrict the possible values of the energy.
In effect,
from Eq.~(\ref{mod_Bessel_small_arg})
and elementary properties of the gamma function~\cite{abr:72},
the small-argument behavior of the modified Bessel
function of the second kind and imaginary index is
\begin{equation}
K_{i\Theta_{l}} (\kappa r)
 \stackrel{(r \rightarrow 0 )}{\sim}
-
\sqrt{ \frac{\pi}{ \Theta_{l}
\sinh  \left( \pi \Theta_{l} \right) } }
\,
\sin
\left[
\Theta_{l} \ln \left( \frac{ \kappa r }{2} \right)
- \delta_{\Theta_{l}}
\right]
\,
\left[ 1 + O \left( r^{2} \right)  \right]
\;  ,
\label{eq:ISP_weak_BS_asymptotic_wf}
\end{equation}
where
\begin{equation}
\delta_{\Theta_{l}}
=
\Im \left[ \ln \Gamma (1+i\Theta_{l}) \right]
\; 
\end{equation}
is the phase of
$\Gamma (1+i\Theta_{l})$.
Then if Eq.~(\ref{eq:ISP_weak_BS_asymptotic_wf}) were
a bound-state wave function,
it would display an oscillatory behavior with
an ever increasing frequency as $r \rightarrow 0$.
In particular,
starting with any finite value of $r$,
the function defined by Eq.~(\ref{eq:ISP_weak_BS_asymptotic_wf})
has an infinite number of zeros,
at the points $r_{n} = 2 \,
\exp \left[ \left( \delta_{\Theta_{l}} - n \pi \right)
/\Theta_{l} \right]/\kappa$
(with $n$ integer),
whence
it represents a state lying above
infinitely  many bound states.
In other words,
the boundary condition has become ineffective as a screening
tool, a situation that  we may describe as a
``loss'' of the boundary condition, and which
permits the existence of a continuum of
bound states extending
from $E=-\infty$ to $E=0$, in agreement with the conclusions
of Ref.~\cite{cam:00a}.
In particular,
 the potential has no ground state,
so that the Hamiltonian is no longer
bounded from below and has lost its self-adjoint
character~\cite{gal:90}.
After due reflection,
this situation makes sense:
the potential has overcome the centrifugal
barrier, a phenomenon that
corresponds to the classical ``fall of the particle to the
center''~\cite{lan:77}.
With regard to the scattering in this strong-coupling regime,
due to the loss of the boundary condition,
it is impossible to determine the phase of the wave function
or relative values of the coefficients
in Eq.~(\ref{eq:ISP_strong_scatt});
in other words, the scattering parameters are ill-defined.

In summary,
for $\lambda<\lambda_{l}^{(\ast)}$, the inverse square potential
is incapable of producing bound states but it scatters
in a manifestly scale-invariant way;
while for $\lambda>\lambda_{l}^{(\ast)}$, it
destroys the discrete character of the bound-state spectrum,
the uniqueness of the scattering solutions, and
the self-adjoint nature of the Hamiltonian.
Here we recognize some of the familiar features
of unregularized transmuting potentials.

The failure of the inverse square potential to provide a
discriminating boundary condition at the origin
for the strong-coupling regime
is now seen as the source of
the  singular behavior required by dimensional transmutation.
In some sense, the key to our regularization procedure
will be the restoration of a sensible boundary condition
for $r=0$.

\subsection{Loss of the Boundary Condition
at the Origin for the Inverse Square Potential}
\label{sec:ISP_loss_BC}

Let us now reexamine the boundary condition at the origin,
which seems to be the most important ingredient in the analysis
of Subsection~\ref{sec:ISP_NR}.
For the inverse square potential,
the standard argument leading to Eq.~(\ref{eq:BC_at_origin})
should be modified with the
replacement~(\ref{eq:ISP_order_replacement_from_free_part}),
so that
\begin{equation}
R_{l}(r) =
\frac{u_{l}(r)}{r^{( D_{0} - 1)/2  }}
 \stackrel{(r \rightarrow 0 )}{\sim}
\mbox{\boldmath\large  $\left\{  \right.$ } \! \! \!
r^{\alpha_{+,l} }
\mbox{\boldmath\large  $,$ }  \!
 r^{\alpha_{-,l} }
\mbox{\boldmath\large  $\left.  \right\}$ } \! \!
\;    ,
\label{eq:ISP_quasi_Laplace_solutions}
\end{equation}
where
\begin{equation}
\alpha_{\pm,l}
= -\nu_{0} \pm s_{l}
\;
\label{ISP_NR_near_origin_alphas}
\end{equation}
are the exponents
arising from the associated indicial equation,
with $s_{l}$ given by
Eq.~(\ref{eq:ISP_order}).
Obviously, Eq.~(\ref{eq:ISP_quasi_Laplace_solutions})
is the small-argument limit of
Eqs.~(\ref{eq:ISP_weak_BS})-(\ref{eq:ISP_strong_scatt}).
However, the power-law functions~(\ref{eq:ISP_quasi_Laplace_solutions})
are no longer solutions of Laplace's equation, so that
the rejection of the second solution, in
the analysis leading to Eq.~(\ref{eq:BC_at_origin}),
is not justified.
In order to clarify this issue it proves convenient
to consider a truncated potential~\cite{lan:77}
\begin{equation}
V_{a}(r) = \left\{ \begin{array} {ll} - \lambda/a^{2}  &  \mbox{for
$r<a$} \\ -\lambda/r^{2}  &  \mbox{for $r\geq a$}
\end{array}
\right.
\label{eq:ISP_truncated}
\;  ,
\end{equation}
which clearly satisfies $
V(r)=\lim_{a \rightarrow 0} V_{a} (r)
$.
The truncated potential
 permits the use of regular boundary conditions for finite
$a$, as a means of deducing
the limiting behavior when
$a \rightarrow 0$.
The solution in the presence
of the truncated potential~(\ref{eq:ISP_truncated}),
for $r$ sufficiently small, is
\begin{equation}
R_{l}(r)
 \stackrel{(r \rightarrow 0,  a \rightarrow 0 )}{\sim}
\left\{ \begin{array} {ll}
 C^{(+)}_{l}
\,
r^{ \alpha_{+,l} }
+  C^{(-)}_{l}
\,
r^{ \alpha_{-,l} }
  &  {\rm for} \; r>a
\\
B_{l} \, r^{l}
 & {\rm for} \; r<a
 \end{array}
\right.
\label{ISP_NR_near_origin}
\;  ,
\end{equation}
where the exponents
$
\alpha_{\pm,l}
$
are given by Eq.~(\ref{ISP_NR_near_origin_alphas}).
Continuity of the logarithmic derivative for $r=a$,
in the limit $a \rightarrow 0$,
 provides the condition
\begin{equation}
\Omega_{l} (a)
=
 \frac{  C^{(+)}_{l} }{  C^{(-)}_{l} }
=
-
\frac{
   \sqrt{ \lambda_{l}^{(\ast)} }
+ \sqrt{ \lambda_{l}^{(\ast)} - \lambda }
}
{
   \sqrt{ \lambda_{l}^{(\ast)}}
- \sqrt{ \lambda_{l}^{(\ast)} - \lambda }
}
\;
 a^{  -2  \sqrt{  \lambda_{l}^{(\ast)}  - \lambda}
}
\;  .
\label{eq:ISP_BC_ratio}
\end{equation}

For the weak-coupling regime,
$\lambda < \lambda_{l}^{(\ast)}$,
the ratio~(\ref{eq:ISP_BC_ratio})
goes to infinity as $a \rightarrow 0$,
and the boundary condition becomes
\begin{equation}
R_{l}(r)
  \stackrel{(r \rightarrow 0 )}{\sim}
 C^{(+)}_{l}
\,
r^{ \alpha_{+,l} }
\; ,
\label{eq:ISP_BC_R}
\end{equation}
so that
\begin{equation}
u_{l}(r)
  \stackrel{(r \rightarrow 0 )}{\sim}
 C^{(+)}_{l}
\,
\sqrt{r}
\,
r^{ s_{l}}
\; .
\label{eq:ISP_BC_u}
\end{equation}
In other words,
the choice is made in favor of the least divergent solution;
then Eq.~(\ref{eq:ISP_BC_u})
implies that the boundary condition for the function $u_{l}(r)$
is~(\ref{eq:BC_at_origin}), just as for regular potentials.
This confirms, a posteriori, that
(\ref{eq:BC_at_origin}) is a valid condition to apply in the
weak-coupling case, as was assumed in the analysis of
Subsection~\ref{sec:ISP_NR},
which led to the proper selection of Bessel functions
in Eqs.~(\ref{eq:ISP_weak_BS}) and (\ref{eq:ISP_weak_scatt}).

However, for the strong-coupling regime,
 $\lambda > \lambda_{l}^{(\ast)}$,
the exponents $\alpha_{\pm} $ are complex conjugate values
and both solutions have the same
behavior.
Then, no criterion can be given to discriminate the ``good''
from the ``bad'' solutions, so that
the  boundary condition at the origin is ``lost.''
Correspondingly,
the ratio of Eq.~(\ref{eq:ISP_BC_ratio}) has no definite
limit for $a \rightarrow 0$, displaying the
same oscillatory behavior
 found in Subsection~\ref{sec:ISP_NR}, i.e.,
\begin{equation}
 \Omega_{l} (a) \propto
\exp \left( -2 i \, \Theta_{l}  \, \ln a \right)
\;  ,
\end{equation}
Then, as  $a \rightarrow 0$, the general solution reduces to
the oscillatory form~(\ref{eq:ISP_weak_BS_asymptotic_wf}).
Notice that Eq.~(\ref{eq:BC_at_origin})
is still satisfied, but it is {\em not\/}
a boundary condition, as it has lost its
discriminating power:
 both solutions are now allowed
(cf.\ Eq.~(\ref{eq:ISP_quasi_Laplace_solutions})).

In short, focusing on the behavior near the origin
exclusively,
we have clarified the meaning
of the loss of boundary condition
and rediscovered the basic features that had already been
predicted in Subsection~\ref{sec:ISP_NR}.

\section{INVERSE SQUARE POTENTIAL: RENORMALIZATION}
\label{sec:ISP_renormalization}

In the first detailed treatment of the inverse square
potential, Ref.~\cite{cas:50},
it was proposed that
an additional orthogonality constraint be imposed on the
eigenfunctions~(\ref{eq:ISP_strong_BS})
in the strong-coupling regime. Effectively, this procedure
restores the discrete nature of the spectrum---a direct application
of Eq.~(\ref{eq:ISP_weak_BS_asymptotic_wf}) gives the energies
$E_{n^{\prime}l}/E_{nl}= e^{2 \pi (n^{\prime} - n)/\Theta_{l}}$.
Nonetheless, the allowed bound-state levels still extend to $-\infty$,
i.e., the Hamiltonian remains unbounded from below.
Subsequently, a number of different
regularization techniques were introduced~\cite{fra:71,par:73,sch:76}
by properly adding regularization parameters of various kinds.
It was soon realized that the strong-coupling Hamiltonian
of the inverse square potential
fails to be self-adjoint, despite it being a symmetric operator;
as its deficiency indices~\cite{gal:90} are $(1,1)$,
its solutions can be regularized with a single
parameter in the form of self-adjoint extensions~\cite{sim:70}.
Alternative attempts simply abandoned self-adjointness
in favor of other requirements or interpretations;
for example, in Ref.~\cite{all:72},
the picture of the ``fall of the particle to the center''
is explicitly implemented by a non-Hermitian condition.

In our approach, we will follow the traditional path
of enforcing the self-adjoint nature of the Hamiltonian---this
time using field-theoretic methods.
Recently, a renormalized solution was presented
along these lines~\cite{gup:93}, which replaced the orthogonality
criterion of Ref.~\cite{cas:50} by a regular boundary condition
at a cutoff point near the origin.
However, this proposal was just limited to the one-dimensional case
and only relied on real-space cutoff regularization.
A second step along this path was the
formulation of the $D$-dimensional generalization of the
cutoff-regularization approach~\cite{cam:00f}.
In this section,
we extend the results of Refs.~\cite{gup:93,cam:00f}
using dimensional regularization and
emphasize those features associated with dimensional transmutation.

\subsection{Dimensional Regularization of the
Inverse Square Potential}
\label{sec:ISP_DR}

Dimensional regularization of the inverse square potential is
done as follows. As discussed in Ref.~\cite{cam:00a},
dimensional analysis implies that
the $D$-dimensional regularization of the inverse square
potential (Eqs.~(\ref{eq:W}) and (\ref{eq:ISP}))
is of the homogeneous form
\begin{equation}
W^{({D})}(r) \propto  r^{-(2 - \epsilon)}
\;  .
\end{equation}
We would like now to confirm this prediction and find the
proportionality coefficient ${\mathcal J}(\epsilon)$.
First, we should find the momentum-space expression
by means of a $D_{0}$-dimensional Fourier transform,
\begin{equation}
\widetilde{W}({\bf k})
= {\mathcal F}_{({D_{0}})} \left\{
W({\bf r}) \right\}
=
\int d^{{D_{0}}} r \; e^{-i{\bf k} \cdot {\bf r}}
\, \frac{1}{r^{2}}
\;  ,
\label{eq:fourier_ISP}
\end{equation}
which we can be computed by means of Bochner's
theorem~\cite{cam:00d}, i.e.,
\begin{equation}
\widetilde{W}({\bf k})
=
\frac{ (2 \pi )^{{D_{0}}/2} }{
    k^{ {D_{0}}/2 -1 }  }
\, \int_{0}^{\infty} dr r^{{D_{0}}/2-2}
J_{{D_{0}}/2-1} (kr)
\;  .
\label{eq:fourier_ISP2}
\end{equation}
Applying the identity~\cite{gra:80}
\begin{equation}
\int_{0}^{\infty}
 x^{\alpha} J_{\beta} (az) dz =
2^{\alpha} a^{-\alpha-1}
\frac{ \Gamma (1/2 + \beta/2 + \alpha/2)}
{ \Gamma (1/2 + \beta/2 - \alpha/2)}
\;  ,
\label{eq:aux_dimcontinuation_ISP}
\end{equation}
which is valid for $- {\rm Re}\, ( \beta) -1 < {\rm Re}\, ( \alpha) < 
1/2$,
the required transform is found to be
\begin{equation}
\widetilde{W}({\bf k})
=
\frac{ (4 \pi )^{{D_{0}}/2} }{
4 k^{ {D_{0}} -2 }  }
\, \Gamma (D_{0}/2 -1)
\;  .
\label{eq:fourier_ISP3}
\end{equation}
Equation~(\ref{eq:fourier_ISP3})
seems to be restricted to $2< {\rm Re} \, (D_{0}) < 5$;
however, the right-hand side is a meromorphic function with
poles at $D_{0}= 2, 0, -2, \ldots$ and is the desired
analytic continuation of the integral.
As we will see next, when the inverse Fourier transform is
applied, the final analytically
continued expression will have no such
restriction.
In effect, applying the
dimensional-continuation prescription
of Ref.~\cite{cam:00a} to $W^{(2)}({\bf r})=1/r^{2}$,
we get
\begin{eqnarray}
W^{({D})}({\bf r}_{{D}})
& = &
\int \frac{
    d^{{D}} k_{
  {D} }} {(2\pi)^{{D}}}
   \; e^{i{\bf k}_{{D}}
\cdot {\bf r}_{{D}} }
\,
\left[ \widetilde{W}({\bf k_{{D_{0}}}})
\right]_{ {\bf k}_{{D_{0}}}
\rightarrow {\bf k}_{{D}} }
\;  ,
\nonumber \\
& = &
\frac{ (4 \pi )^{{D_{0}}/2} }{ 4  }
\, \Gamma \left( D_{0}/2 -1 \right) \,
\int \frac{ d^{{D}} k_{ {D} }}
{(2\pi)^{{D}}}
   \; e^{i{\bf k}_{{D}}
\cdot {\bf r}_{{D}} }
\,
\left[  k_{{D_{0}}}^{-{D_{0}}+2}
\right]_{ {\bf k}_{{D_{0}}}
 \rightarrow {\bf k}_{{D}} }
\nonumber \\
& = &
\frac{ (2 \pi )^{\epsilon/2} }{  r^{ {D}/2 -1 }  }
\, 2^{{D_{0}}/2-2}
\, \Gamma \left( D_{0}/2 -1  \right)
\,
\int_{0}^{\infty} dk \,  k^{{D}/2
-{D_{0}} +2}
J_{{D}/2 -1} (kr)
\;  ,
\label{eq:dimcontinuation_ISP}
\end{eqnarray}
which, by means of Eq.~(\ref{eq:aux_dimcontinuation_ISP}),
amounts to
\begin{equation}
W^{({D_{0}} - \epsilon)}(r) =
\pi^{\epsilon/2} \,
\Gamma \left(1 - \frac{\epsilon}
{2} \right)
\, r^{-(2 - \epsilon)}
\; .
\label{eq:dimcontinuation_ISP2}
\end{equation}
Even though the last integration is
restricted to $3-D_{0}< {\rm Re} (D_{0}-D ) <2$,
analytic continuation of the right-hand of
Eq.~(\ref{eq:dimcontinuation_ISP2}) allows us to extend its validity
for arbitrary values of $\epsilon \neq -2, -4, \dots$
(notice that the previous restriction has been lifted);
this is perfectly fine anyway,  because all that is needed is
the limit $\epsilon= 0^{+}$.

Then, for the dimensionally
regularized problem,
using the notation
\begin{equation}
{\mathcal J}(\epsilon) = \pi^{\epsilon/2}
\Gamma \left( 1 -
\frac{\epsilon}{2} \right)
\; ,
\label{eq:c_eps}
\end{equation}
we have the $D$-dimensional Schr\"{o}dinger equation
\begin{equation}
\left[
\nabla^{2}_{{D}} +  E  + \frac
{\lambda \, \mu^{\epsilon}
{\mathcal J}(\epsilon)}{r^{2 -
\epsilon}}
\right]
\Psi({\bf r}) = 0
\;  ,
\label{eq:ISPSchr}
\end{equation}
which, for the state of angular momentum $l$,
reads
\begin{equation}
\left\{
\frac{ d^{2}}{dr^{2} } + E
 +
\frac{\lambda \, \mu^{\epsilon}
{\mathcal J}(\epsilon)}{r^{2-\epsilon}}
-
\frac{ \left[ l+ D (\epsilon) / 2 - 1 \right]^{2} -
1/4}{r^{2}}
\right\}
u_{l}(r) = 0
\;  .
\label{eq:ISP_schr}
\end{equation}

We are now ready to solve Eq.~(\ref{eq:ISP_schr}). As it stands, it
does not look familiar when $\epsilon$ is not an integer.
However, it can be conveniently transformed into
an asymptotically soluble problem by means of a
duality transformation~\cite{qui:79},
whose properties we study in Appendix~\ref{sec:powerlaw}.
For the particular case at hand,
applying the transformation
(cf.\ (\ref{eq:transf_Schr4}))
\begin{equation}
\left\{ \begin{array}{l}
\left| E \right|^{1/2} r
= z^{ 2/\epsilon}
\\
|E|^{-D/4} \, u (r)= w (z) \, z^{ 1/\epsilon -
 1/2 }
\end{array} \right.
\;  ,
\end{equation}
the dual of Eq.~(\ref{eq:ISP_schr}) becomes
 Eq.~(\ref{eq:radialSchr_ISP}), which we write in the condensed form
\begin{equation}
\left\{
\frac{d^{2}}{dz^{2}} +
\widetilde{\eta} -
\widetilde{\mathcal V}_{\epsilon}(z)
- \frac{ [p_{l}(\epsilon)]^{2} - 1/4 }
{z^{2}}
\right\}
w_{l,\epsilon}
(z)
= 0
\;  ,
\label{eq:ISPSchr2}
\end{equation}
with
a dual dimensionless energy
\begin{equation}
\widetilde{\eta} = \frac{4 \lambda \, \mu^{\epsilon}
\, {\mathcal J}(\epsilon)}{\epsilon^{2}}
\,
|E|^{-\epsilon/2 }
\;  ,
\label{eq:dual_energy_ISP}
\end{equation}
 a dual
potential energy  term,
\begin{equation}
\widetilde{\mathcal V}_{\epsilon} (z) = - \sigma \, 
\frac{4}{\epsilon^{2}}
z^{4/\epsilon -2}
\;  ,
\label{eq:dual_PE_ISP}
\end{equation}
and
a dual angular momentum  variable
\begin{equation}
p_{l}(\epsilon) =
\frac{2}{\epsilon}
\left[l+\frac{D(\epsilon)}{2} - 1 \right]
\;  .
\label{eq:ISP_p_l_epsilon}
\end{equation}
In our subsequent analysis, we will
often rewrite Eq.~(\ref{eq:ISP_p_l_epsilon}) in the form
\begin{equation}
p = p_{l}(\epsilon) =
\frac{2}{\epsilon}
\,
\left(
\lambda_{l}^{(\ast)}
\right)^{1/2}
-1
=
\frac{2}{\epsilon} \,
\left(
\lambda_{l}^{(\ast)}
\right)^{1/2}
  \,
\left[ 1 - \frac
{\epsilon}{2}
\left(
\lambda_{l}^{(\ast)}
\right)^{-1/2}
\right]
\; ,
\label{eq:ISP_p_l_epsilon2}
\end{equation}
in terms of the critical coupling~(\ref{eq:ISP_critical_coupling}).
Equation~(\ref{eq:ISP_p_l_epsilon2}) provides an alternative expansion
 parameter $p$, such that, as $\epsilon \rightarrow 0$,
$p \rightarrow \infty$.
It should be noticed in passing that Eq.~(\ref{eq:dual_energy_ISP})
already provides the functional form~(\ref{eq:eigenvalues_general2})
required for any possible eigenvalue equation of the inverse
square potential, with
\begin{equation}
\Xi (\epsilon)
=
 \frac{4
\, {\mathcal J}(\epsilon)}{\epsilon^{2}}
\,
\widetilde{\eta}^{-1}
\;  .
\label{eq:ISP_eigenvalues}
\end{equation}

As usual, as the problem is now regularized,
Eq.~(\ref{eq:ISPSchr2})
should be solved with the regular
boundary condition at the origin,
Eq.~(\ref{eq:BC_at_origin}), so that
\begin{equation}
w_{l,\epsilon}
(0)
=
0
\;  .
\label{eq:bc_w_at_0}
\end{equation}

Even though the solutions to
Eq.~(\ref{eq:ISPSchr2}) cannot be expressed
in terms of any standard special functions
for arbitrary $\epsilon>0$,
it is still possible to
write their asymptotic form with respect to $\epsilon \rightarrow 0$,
in terms of Bessel functions, as we show below.
The key to this asymptotic analysis lies in the singular
limiting form of the transformed potential~(\ref{eq:dual_PE_ISP}),
namely,
\begin{equation}
\lim_{\epsilon \rightarrow 0}
\widetilde{\mathcal V}_{\epsilon} (z)
=
\left\{ \begin{array}{ll}
0
&
{\rm for} \; z<1
\\
- \sigma \infty
&
{\rm for} \; z>1
\end{array} \right.
\;  ;
\label{eq:limiting_dual_PE_ISP}
\end{equation}
in particular,
for $E<0$ (i.e., $\sigma=-1$),
$\widetilde{\mathcal V}_{\epsilon} (z) $
behaves as an infinite
 hyperspherical potential well.
Because of the singular nature of the
limit~(\ref{eq:limiting_dual_PE_ISP}),
an asymptotically exact hierarchy emerges, whereby
Eq.~(\ref{eq:ISPSchr2})
is split into two distinct differential equations:
the first one, without the term $\widetilde{\mathcal V}_{\epsilon} (z) 
$,
for the interior region $ z<z_{2} \approx 1$,
\begin{equation}
\left\{
\frac{d^{2}}{dz^{2}} +
\widetilde{\eta}
- \frac{ [p_{l}(\epsilon)]^{2} - 1/4 }
{z^{2}}
\right\}
w^{(<)}_{l,\epsilon}
(z)
=
0
\;  ;
\label{eq:ISPSchr2_int}
\end{equation}
and the second one,
without the terms proportional to $1/z^{2}$,
for the  exterior region $ z>z_{2} \approx 1$,
\begin{equation}
\left[
\frac{d^{2}}{dz^{2}} -
\widetilde{\mathcal V}_{\epsilon}(z)
\right]
w^{(>)}_{l,\epsilon} (z)
=
0
\;  .
\label{eq:ISPSchr2_ext}
\end{equation}

Equation~(\ref{eq:ISPSchr2_int})
 is a Bessel differential equation of order
 $p=p_{l}(\epsilon)$, whose regular
solutions that satisfy the boundary condition~(\ref{eq:bc_w_at_0})
are of the form
\begin{equation}
w^{(<)}_{l,\epsilon}
(z)
=
B_{l}
\,
\sqrt{z}
\,
J_{p}( \widetilde{\eta}^{1/2} z)
\; .
\label{eq:ISP_w_int}
\end{equation}
On the other hand,
Eq.~(\ref{eq:ISPSchr2_ext})
can be taken to a standard Bessel form by another duality
transformation; it is easy to verify that
\begin{equation}
w^{(>)}_{l,\epsilon}
(z)
=
\sqrt{z}
\,
{\mathcal C}_{\epsilon/4}(  \sqrt{\sigma} z^{2/\epsilon})
\; ,
\label{eq:ISP_w_ext1}
\end{equation}
where ${\mathcal C}_{\epsilon/4}$
stands for a linear combination of ordinary
Bessel functions
of order $\epsilon/4$; in fact, more explicitly,
\begin{equation}
w^{(>)}_{l,\epsilon}
 (z)
=
\left\{ \begin{array}{ll}
A_{l}  \,
\sqrt{z}
\,
K_{\epsilon/4} (  z^{2/\epsilon})
 &
{\rm for} \; \sigma=-1
\\
\sqrt{z}
\,
\left[
\widetilde{A}_{l}^{(+)} H^{(1)}_{\epsilon/4} (  z^{2/\epsilon})
+
\widetilde{A}_{l}^{(-)} H^{(2)}_{\epsilon/4} (  z^{2/\epsilon})
\right]
&
 {\rm for} \; \sigma=1
\end{array} \right.
\;  ,
\label{eq:ISP_w_ext2}
\end{equation}
where the first line
includes a modified Bessel function of the second kind,
whereas the second line includes a linear combination
of Hankel functions.
Their physical meaning becomes transparent when
transformed back from
the dual to the original space,
in which
\begin{equation}
u_{l}(r)
 \stackrel{(r \rightarrow \infty, \epsilon \rightarrow 0)}{\sim}
\left\{ \begin{array}{ll}
A_{l}  \,
\sqrt{r} \,
 K_{0} (  \kappa r )
 &
{\rm for} \; E=-\kappa^{2}<0
\\
\sqrt{r} \,
\left[
\widetilde{A}_{l}^{(+)} H^{(1)}_{0} (  k r )
+
\widetilde{A}_{l}^{(-)} H^{(2)}_{0} (  k r )
\right]
&
 {\rm for} \; E=k^{2}>0
\end{array} \right.
\;  .
\label{eq:ISP_w_ext3}
\end{equation}
Equation~(\ref{eq:ISP_w_ext3})
represents the asymptotic behavior of the wave function,
when $r$ is large, for the bound-state  ($E<0$)
and scattering  ($E>0$) sectors of the inverse
square potential. Of course,
the asymptotic behavior of Eq.~(\ref{eq:ISP_w_ext3})
justifies a posteriori
the choice of Bessel functions in
Eq.~(\ref{eq:ISP_w_ext2}).

The boundary point $z_{2}$
separating the two regimes described by
Eqs.~(\ref{eq:ISPSchr2_int})
and (\ref{eq:ISPSchr2_ext})
 is infinitesimally close to $z=1$,
but further precision is required.
In effect, the critical separation
 takes place at the point $z_{2}$ where
\begin{equation}
\left|
\widetilde{\eta} -
\frac{ [p_{l}(\epsilon)]^{2} - 1/4 }
{z_{2}^{2}}
\right|
=
\left|
\widetilde{\mathcal V}_{\epsilon}(z_{2})
\right|
\;  ,
\label{eq:z_2_defined}
\end{equation}
because,
away from
it, the term $\widetilde{\mathcal V}_{\epsilon}(z_{2}) $
will abruptly become negligible for $z<z_{2}$
and will overwhelmingly dominate over the other terms
for $z>z_{2}$.
From Eqs.~(\ref{eq:dual_energy_ISP})--(\ref{eq:ISP_p_l_epsilon2}),
it follows that the condition~(\ref{eq:z_2_defined})
amounts to
\begin{equation}
p^{2} \left( \frac{\lambda}{\lambda_{l}^{(\ast)}} -1
\right) \left[ 1 +
O \left( \frac{1}{p} \right)
\right]
=
\frac{p^{2}}{ \lambda_{l}^{(\ast)}}
\,
z_{2}^{4/\epsilon}
\,
\left[ 1 + O \left( \frac{1}{p} \right) \right]
\;  ,
\end{equation}
so that
\begin{equation}
z_{2}= \Theta_{l}^{\epsilon/2}
\;  ,
\label{eq:z_2}
\end{equation}
where $\Theta_{l}$ is given by Eq.~(\ref{eq:Theta}).
With the value
of $z_{2}$ so determined, the solutions to
Eqs.~(\ref{eq:ISPSchr2_int})
and (\ref{eq:ISPSchr2_ext})
should be matched at  $z=z_{2}$
through the logarithmic derivative
\begin{equation}
{\mathcal L} (\Theta_{l})
 =
\left.
z \,
\frac{d \ln \left( w/\sqrt{z}  \right) }{dz} \right|_{z=z_{2}}
\;  .
\label{eq:log_derivative}
\end{equation}
For the interior solution~(\ref{eq:ISP_w_int}),
the logarithmic derivative  takes the form
\begin{equation}
{\mathcal L}^{(<)} (\Theta_{l})
 =
\left.
y
\,
\frac{d \ln J_{p}(y) }
{dy}
\right|_{y=\widetilde{\eta}^{1/2}z_{2}}
\;  ,
\label{eq:log_derivative_int}
\end{equation}
whereas, for the exterior solution~(\ref{eq:ISP_w_int}),
it becomes
\begin{equation}
{\mathcal L}^{(>)} (\Theta_{l})
 =
\frac{2}{\epsilon}
\,
\Theta_{l}
\,
\frac{d \ln {\mathcal C}_{\epsilon/4}(\Theta_{l})}{d \Theta_{l}}
\;  .
\label{eq:log_derivative_ext}
\end{equation}

Having developed the general framework for
the analysis of the inverse square potential in terms of its
dual problem,
we now turn to the distinct
details of the bound-state and scattering sectors.

\subsection{Bound-State Sector for the Inverse Square Potential}
\label{sec:ISP_bound_states}

The bound
states of the inverse square potential (if any)
should be characterized by the energy condition $E<0$,
which essentially converts the dual potential
into an infinite hyperspherical well
in the limit $\epsilon \rightarrow 0^{+}$, as displayed by
Eq.~(\ref{eq:limiting_dual_PE_ISP}).
This suggests that, as a first approximation,
the boundary condition at $z_{2}$ may be replaced by
\begin{equation}
w^{(<)}_{l,\epsilon}
(1)
\approx 0
\;  ,
\label{eq:bc_w_at_1}
\end{equation}
whence
Eq.~(\ref{eq:ISP_w_int})
immediately gives the spectrum through
the condition
\begin{equation}
J_{p}(\widetilde{\eta}^{1/2}) \approx
 0
\; ,
\end{equation}
from which the corresponding eigenvalues are
\begin{equation}
\widetilde{\eta}_{n_{r}}
\approx
\left( j_{p,n_{r}} \right)^{2}
\;  ,
\label{eq:eta_tilde_n_approx}
\end{equation}
where $j_{p,n_{r}}$ is the $n_{r}$th zero of the Bessel
function of order $p$.
Therefore, 
from Eq.~(\ref{eq:dual_energy_ISP}),
the regularized eigenvalue condition
becomes
\begin{equation}
\lambda
\,
\left[\frac{4{\mathcal J}(\epsilon)(j_{p,n_{r}})^{-2}}{\epsilon^{2}}
\right]
\left(
\frac{|E|}{ \mu^{2} }
\right)^{- \epsilon/2 }
\approx 1
\;  ,
\label{eq:ISP_eigenvalues2}
\end{equation}
where $n_{r}$ is now confirmed
as the radial quantum number---the ordinal number
for the stationary radial wave functions.
Equation~(\ref{eq:ISP_eigenvalues2})
is of the form of the master
eigenvalue equation~(\ref{eq:eigenvalues_general2}),
with an energy generating function
\begin{equation}
\Xi_{n_{r} l}(\epsilon) \approx
\frac{4{\mathcal J}(\epsilon)}{\epsilon^{2}}
\left( j_{p,n_{r}} \right)^{-2}
\;  .
\label{eq:F_eps}
\end{equation}

Even though the remarks of the previous paragraph are
essentially correct, for a full comparison with the scattering
sector of the theory, it is necessary to evaluate
the logarithmic derivatives~(\ref{eq:log_derivative_int})
 and (\ref{eq:log_derivative_ext}),
which will give extra terms in the energy expressions.
Therefore, we need to first analyze the asymptotic behavior of
the interior solution~(\ref{eq:ISP_w_int})
with respect to $p \rightarrow \infty$,
and then find the proper values of the logarithmic derivatives.

The limit $\epsilon \rightarrow 0$
of Eq.~(\ref{eq:ISP_w_int})
relies on well-known properties of the
Bessel functions of large order~\cite{abr:72}.
We will start with Debye's asymptotic expansion,
\begin{equation}
J_{p} ( p\sec \beta)
 \stackrel{(p \rightarrow \infty)}{\sim}
\sqrt{ \frac{2}{\pi p \tan \beta} }
\left\{
\cos
\left[
p \tan \beta - p \beta - \frac{\pi}{4}
\right]
+ O( p^{-1} )
\right\}
\;  ,
\label{eq:Debye_expansion}
\end{equation}
where the argument
will become
\begin{equation}
y = p \sec \beta = \widetilde{\eta}^{1/2} z_{2}
\;  .
\label{eq:Debye_variable}
\end{equation}
For our analysis,
a few relations between these variables are in order;
in particular,
\begin{equation}
y
=
p
\,
\sqrt{ \frac{\lambda}{ \lambda_{l}^{(\ast)}} }
\,
\left[ 1 + O \left( \epsilon \right)
\right]
=
p \,
\sqrt{1 +  \frac{\Theta_{l}^{2} }{ \lambda_{l}^{(\ast)} }  }
\,
\left[ 1 + O \left( \epsilon \right)
\right]
\;  ,
\end{equation}
and
\begin{equation}
\tan \beta =
\sqrt{ \left(y/p\right)^{2} - 1 }
=
\frac{\Theta_{l} } { \left( \lambda_{l}^{(\ast)} \right)^{1/2} }
\;  .
\end{equation}
Then the logarithmic derivative~(\ref{eq:log_derivative_int})
becomes
\begin{equation}
{\mathcal L}^{(<)} (\Theta_{l})
 =
-p
\,
\frac{\Theta_{l}}{  \left( \lambda_{l}^{(\ast)} \right)^{1/2} }
\tan
\left\{
p \left[
\frac{\Theta_{l}}{  \left( \lambda_{l}^{(\ast)} \right)^{1/2} }
-
\tan^{-1}
\left( \frac{\Theta_{l}}{  \left( \lambda_{l}^{(\ast)} \right)^{1/2} } 
\right)
\right]
-
\frac{\pi}{4}
\right\}
+ O(1)
\;  .
\label{eq:BS_log_derivative_int1}
\end{equation}

Equation~(\ref{eq:BS_log_derivative_int1})
displays an anomalous behavior as $p \rightarrow \infty$
in that the tangent function will oscillate wildly
from $-\infty$ to $\infty$,
thus rendering the regularized problem
ill-defined.
The cure for this behavior is afforded by the
renormalization of the coupling constant $\lambda=\lambda (\epsilon)$,
which implies a corresponding renormalization
of $\Theta_{l}= \Theta_{l} (\epsilon)$ (from Eq.~(\ref{eq:Theta})),
in such a way that $\Theta_{l} \rightarrow 0$
as $\epsilon \rightarrow 0$.
Then this amounts to the realization of the limiting critical coupling
$\lambda \rightarrow \lambda_{l}^{(\ast)} + 0^{+}$.
A more careful analysis of this limiting behavior is afforded by
\begin{equation}
{\mathcal F} =
\tan \beta - \beta
=\frac{\beta^{3}}{3} + O(\beta^{5})
=
\frac{\Theta_{l}^{3}}{3  \left( \lambda_{l}^{(\ast)} \right)^{3/2} } +
O(\Theta_{l}^{5})
\;  ,
\label{Debye_F}
\end{equation}
so that the finiteness of the argument of the tangent implies
that $p \Theta_{l}^{3}$ should be finite; then
\begin{equation}
\Theta_{l} =
O
\left(p^{-1/3}\right)
\;
\label{eq:order_of_Theta}
\end{equation}
and
\begin{equation}
y = p +
p^{1/3}  x
\;  ,
\label{eq:order_of_y}
\end{equation}
where $x$ is a finite variable.

On the other hand,
using the small-argument behavior of the
modified Bessel function of the second kind~\cite{abr:72},
Eq.~(\ref{mod_Bessel_small_arg}),
it follows that
\begin{equation}
  K_{\epsilon/4} (z^{2/\epsilon})
  \stackrel{(z \rightarrow 0)}{\sim}
\left[
\frac{2}{\epsilon}
\left(
\frac{1}{ \sqrt{z} } - \sqrt{z}  \right)
- \frac{1}{2} \,
\left( \gamma - \ln 2 \right)
\left(
\frac{1}{ \sqrt{z} } + \sqrt{z}  \right)
 \right]
\,
\left[ 1 + O(\epsilon^{2},z^{4/\epsilon}) \right]
\;  ,
\label{rescaled_mod_Bessel_small_arg}
\end{equation}
so that, from Eqs.~(\ref{eq:z_2}),
(\ref{eq:log_derivative_ext}),
and
(\ref{eq:order_of_Theta}), the logarithmic derivative becomes
\begin{equation}
{\mathcal L}^{(>)} (\Theta_{l})
 =
\frac{2}{\epsilon}
\,
\left( \ln \Theta_{l} -c \right)^{-1}
\,
\left[ 1 + O(\epsilon^{2/3} )
 \right]
\;  ,
\label{eq:BS_log_derivative_ext}
\end{equation}
where
\begin{equation}
c= \ln 2 - \gamma
\;  .
\label{eq:c_log_derivative}
\end{equation}
Equation~(\ref{eq:BS_log_derivative_ext})
implies that, as $\epsilon \rightarrow 0$,
\begin{equation}
\frac{ w^{(>)}
(z_{2}) }{  \sqrt{z_{2}} }
 \stackrel{(\epsilon \rightarrow 0)}{\sim}
\frac{\epsilon}{2}
\,
\left( \ln \Theta_{l} -c \right)
\,
\left.
\frac{d}{dz}
\left[
\frac{  w^{(>)}
(z) }{  \sqrt{z} }
\right]
\right|_{z=z_{2}}
\;  ,
\end{equation}
so that continuity of the wave function implies that
Eq.~(\ref{eq:bc_w_at_1}) is indeed correct to zeroth order.

 The interior logarithmic derivative can now be evaluated
from the asymptotic expansion~\cite{abr:72}
\begin{equation}
 J_{p} \left( p +
 p^{1/3}  x \right)
 \stackrel{(p \rightarrow \infty)}{\sim}
\frac{2^{1/3}}{p^{1/3}}
\,
Ai \left( - 2^{1/3} x \right)
+  O \left(  p^{-1}  \right)
\;  ,
\label{eq:large_order_Bessel_from_Airy}
\end{equation}
where $Ai $ is the Airy function of the first kind~\cite{abr:72}.
The  zeroth order approximation
to the logarithmic derivative
amounts to $w(1)=0$ or ${\mathcal L}^{(<)} =\infty$,
so that the value of $x$, to this order, is provided
by the zeros of the Airy function,
\begin{equation}
Ai \left( - 2^{1/3} x^{(0)}_{n_{r}} \right) = 0
\; .
\label{eq:Airy_zeros}
\end{equation}
More precisely, if $a_{n_{r}}$ is the $n_{r}$th negative zero of
$Ai$,
then
\begin{equation}
C_{n_{r}} = x^{(0)}_{n_{r}}
=
2^{-1/3} a_{n_{r}}
\;  ,
\label{eq:C_n_large_order_Bessel}
\end{equation}
where
we have identified the leading contribution in the
asymptotic formula for the zeros
of the Bessel function of
large order~\cite{jah:45},
\begin{equation}
 j_{p,n_{r}} = p + C_{n_{r}} \, p^{1/3} + O \left(p^{-2/3} \right)
\;  ;
\label{eq:j_pn}
\end{equation}
for example,
$C_{1}=1.8558$,
$C_{2} = 3.2447$, etc.
It should be pointed out that,
from the asymptotic form of the Airy functions, which amounts to
a WKB integration of Eq.~(\ref{eq:ISPSchr2_int}),
an  approximate---but extremely accurate---formula
can be derived for these coefficients, namely~\cite{cam:00f},
\begin{equation}
C_{n_{r}} \approx
\frac{ \left( 3 \pi  \right)^{2/3}   }{2}   \,
\left( n_{r} - \frac{1}{4}
\right)^{2/3}
\; .
\label{eq:C_n_large_order_Bessel_approx}
\end{equation}
Then, for the next order,
\begin{equation}
x_{n_{r}} = C_{n_{r}}
+ \delta_{n_{r}}
\;  ,
\end{equation}
with $\delta_{n_{r}} \ll C_{n_{r}}  $,
one can expand the Airy function in a Taylor series
in the neighborhood
of $ x^{(0)}_{n_{r}} $, so that  
Eq.~(\ref{eq:log_derivative_int}) becomes
\begin{eqnarray}
{\mathcal L}^{(<)} (\Theta_{l})
& =  &
- 2^{1/3} \,  p^{2/3}
\,
\frac{     Ai^{\, \prime} \left( - 2^{1/3} x \right) }
{ Ai \left( - 2^{1/3} x \right) }
\,
\left[
1 + O
\left(
p^{-2/3} \right) \right]
\nonumber \\
& = &
\frac{  p^{2/3}  }{\delta_{n_{r}}  }
\,
\left[
1 + O \left( \delta_{n_{r}}, p^{-2/3} \right) \right]
\;  .
\label{eq:BS_log_derivative_int2}
\end{eqnarray}
The equality of the logarithmic derivatives,
Eqs.~(\ref{eq:BS_log_derivative_ext})
and (\ref{eq:BS_log_derivative_int2}),
gives
\begin{equation}
\delta_{n_{r}}
=
p^{-1/3}
\left(
\lambda_{l}^{(\ast)} \right)^{1/2}
\,
\left[
-c + \ln \Theta_{l} \right]
\,
\left[
1 +
 O \left( \epsilon^{1/3} \right) \right]
\; ,
\label{eq:delta_n}
\end{equation}
whence
\begin{equation}
y_{n_{r}}
=
p
\left\{
1 +
C_{n_{r}}
\,
p^{-2/3}
+
\left( \lambda_{l}^{(\ast)} \right)^{1/2}
\,
\left[
-c + \ln \Theta_{l} \right]
p^{-1}
+
O
\left( p^{-4/3} \right)
\right\}
\;  .
\label{eq:y_n}
\end{equation}
Therefore,
from Eqs.~(\ref{eq:z_2}) and (\ref{eq:Debye_variable}),
the correct replacement for
Eq.~(\ref{eq:eta_tilde_n_approx}) is
\begin{equation}
\widetilde{\eta}_{n_{r}}
= \left(
\frac{y}{z_{2}}
\right)^{2}
=
p^{2}
\,
\left\{
1 + 2 \, C_{n_{r}} \, p^{-2/3}
- 2
\left( \lambda_{l}^{(\ast)} \right)^{1/2}
\,
c
\,
p^{-1}
+
O
\left( p^{-4/3} \right)
\right\}
\;  .
\label{eq:eta_tilde_n}
\end{equation}
Finally,
from Eq.~(\ref{eq:c_eps}),
the  expansion of $  {\mathcal J}(\epsilon)$
about $\epsilon=0^{+}$ is
\begin{equation}
  {\mathcal J}(\epsilon)
= 1 + \frac{\epsilon}{2} \left( \gamma + \ln \pi \right)
+ O(\epsilon^{2} )
\;  ,
\label{eq:c_epsilon}
\end{equation}
which combined with Eqs.~(\ref{eq:ISP_p_l_epsilon2})
and (\ref{eq:eta_tilde_n}),
gives the asymptotically exact
expression for the eigenvalue function of
Eqs.~(\ref{eq:eigenvalues_general2}) and (\ref{eq:ISP_eigenvalues}),
namely,
\begin{eqnarray}
\Xi_{n_{r}l} (\epsilon)
& =  &
\frac{1}{\lambda_{l}^{(\ast)}}
\,
\left\{
1 -2 p^{-2/3}
C_{n_{r}} +
p^{-1} \left(
 \lambda_{l}^{(\ast)} \right)^{1/2}
\left[
\ln 4 \pi
- \gamma
+  2 \, \left( \lambda_{l}^{(\ast)} \right)^{-1/2}
\right]
+  O \left( p^{-4/3} \right)
\right\}
\\
& = &
 \frac{1}{\lambda_{l}^{(\ast)}}
\left\{
1 - 2^{1/3} C_{n_{r}} \,
(\lambda_{l}^{(\ast)})^{-1/3}\epsilon^{2/3}+\frac{\epsilon}{2}
\left[
 \ln  4 \pi - \gamma
+2 (\lambda_{l}^{(\ast)})^{-1/2}  \right]
+  O \left( \epsilon^{4/3} \right)
\right\}
\;  ,
\label{eq:ISP_eigenvalues3}
\end{eqnarray}
which is of the form~\cite{cam:00a}
\begin{equation}
\Xi_{n}(\epsilon)
=
\left[ L_{n}  (\epsilon) \right]^{-1}
\,
\left[ 1 + \frac{\epsilon}{2}  \,
{\mathcal G}_{n} (\epsilon)   \right]
\;  ,
\label{eq:master_eigenvalue_function_practical_expansion}
\end{equation}
with a constant critical coupling function
\begin{equation}
L_{n_{r} \, l}
=
\lambda_{l}^{(\ast)}
= \left( l + \frac{D_{0}}{2}
- 1 \right)^{2}
\;
\label{eq:ISP_critical_coupling_DT}
\end{equation}
(independent of $n_{r}$),
and with
\begin{equation}
{\mathcal G}_{n_{r} \, l}(\epsilon)=
-2^{4/3}C_{n_{r}} \, (\lambda_{l}^{(\ast)})^{-1/3}
\epsilon^{-1/3}
+
\left[ \ln 4 \pi - \gamma
+2 (\lambda_{l}^{(\ast)})^{-1/2} \right] + O (\epsilon^{1/3})
\;  .
\label{eq:ISP_G_function}
\end{equation}
From Eq.~(\ref{eq:ISP_G_function}),
we  see that
\begin{equation}
 {\mathcal G}_{_{\rm (gs)}}
 (\epsilon)
=
-2^{4/3} \, C_{1} \, (\lambda^{(\ast)}_{_{\rm (gs)}} )^{-1/3}
\epsilon^{-1/3}
+
\left[ \ln 4 \pi - \gamma
+2 (\lambda_{_{\rm (gs)}}^{(\ast)})^{-1/2} \right] + O (\epsilon^{1/3})
\;  ,
\label{eq:ISP_G_function_GS}
\end{equation}
where
\begin{equation}
\lambda_{_{\rm (gs)}}^{(\ast)} = \lambda_{0}^{(\ast)} = \nu_{0}^{2}
=
\left( D_{0}/2  -  1 \right)^{2}
\;
\label{eq:principal_lambda_star}
\end{equation}
is the ``principal coupling constant,'' i.e.,  the critical
coupling for the ground state ($l=0$).

Equations~(\ref{eq:eigenvalues_general2}) and
(\ref{eq:master_eigenvalue_function_practical_expansion})--(\ref{eq:ISP_G_function})
provide the regularized energies
\begin{eqnarray}
|E_{n_{r} l}|
& = &
\mu^{2} \,
\left(
\frac{ \lambda}{  \lambda_{l}^{(\ast)} }
\right)^{2/\epsilon}
\,
\exp
\left[
{\mathcal G}_{n_{r} \, l}(\epsilon)
\right]
\nonumber \\
& = &
\mu^{2} \,
\left(
\frac{ \lambda}{  \lambda_{l}^{(\ast)} }
\right)^{2/\epsilon}
\,
\exp
\left\{
-2^{4/3}  C_{n_{r}}
\left( \lambda_{l}^{(\ast)}  \right)^{-1/3}
\epsilon^{-1/3}
+ \left[ \ln 4 \pi - \gamma  + 2
\left( \lambda^{(\ast)}_{l} \right)^{-1/2}
\right]
\right\}
\;  .
\label{eq:ISP_DR_regularized_energies}
\end{eqnarray}
Ground-state renormalization can be implemented from
Eq.~(\ref{eq:ISP_DR_regularized_energies}), by means of the
regularized coupling~\cite{cam:00a}
 \begin{equation}
\lambda(\epsilon)
=
\left[
\,
\Xi_{_{\rm (gs)}}  (\epsilon)
\right]^{-1}
\,
\left[ 1 +
\frac{\epsilon}{2}
\, g^{(0)}
+ o(\epsilon)  \right]
\;  ,
\label{eq:epsilon_dependence_coupling}
\end{equation}
which now becomes
\begin{equation}
\lambda (\epsilon)
=
\lambda_{_{\rm (gs)}}^{(\ast)}
\left\{
1 +
2^{1/3}
\left( \lambda_{_{\rm (gs)}}^{(\ast)}
 \right)^{-1/3} \epsilon^{2/3}
C_{1}
+ \frac{\epsilon}{2}
\left[ g^{(0)}
-
\left(  \ln 4 \pi - \gamma \right)
-2 (\lambda_{_{\rm (gs)}}^{(\ast)})^{-1/2}
\right]
\right\}
+
o(\epsilon)
\;  ,
\label{ISP_coupling_reg}
\end{equation}
with an arbitrary finite part $g^{(0)}$,
so that, for the ground state,
\begin{equation}
E_{_{\rm (gs)}}
=
- \mu^{2}
\exp \left[
g^{(0)}
\right]
\leadsto
- \mu^{2}
\;
\label{eq:ISP_GS}
\end{equation}
(cf.\ Eq.~(\ref{delta_gs})),
while for any regularized bound state
\begin{eqnarray}
|E_{n_{r} l}|
 =
\mu^{2} \,
\left(
\frac{ \lambda_{_{\rm (gs)}}
}{  \lambda_{l}^{(\ast)} }
\right)^{2/\epsilon}
\,
\exp
\left\{  
 - 2^{4/3} 
 \left[ C_{n_{r}} 
\left( \lambda_{l}^{(\ast)}  \right)^{-1/3}
\right.
\right.
& - & 
\left.
C_{1}  
\left( \lambda_{_{\rm (gs)}}^{(\ast)}
 \right)^{-1/3}
\right]
\epsilon^{-1/3}
\nonumber
\\
& + &
\left.
\! 
g^{(0)}
+
2
\left[ 
\left( \lambda_{l}^{(\ast)}  \right)^{-1/2}
-
\left( \lambda_{_{\rm (gs)}}^{(\ast)}
 \right)^{-1/2}
\right]
\right\}
\;  .
\label{eq:ISP_DR_regularized_energies2}
\end{eqnarray}
Equation~(\ref{ISP_coupling_reg})
explicitly shows that
\begin{equation}
\lambda (\epsilon)
 \stackrel{(\epsilon \rightarrow 0 )}{\longrightarrow}
  \lambda_{_{\rm (gs)}}^{(\ast)} + 0^{+}
\; ;
\label{eq:ISP_renormalized_coupling}
\end{equation}
in other words, when the system is renormalized,
the coupling becomes critically strong with respect to the ground
state (which, as we will see below, has $l=0$).

Equations~(\ref{eq:ISP_eigenvalues3})--(\ref{eq:ISP_renormalized_coupling})
are easily interpreted.
As expected, the critical coupling defined
by Eq.~(\ref{eq:ISP_critical_coupling})
from the unregularized theory
becomes the critical coupling for the regularized theory,
Eq.~(\ref{eq:ISP_critical_coupling_DT}),
with respect to dimensional transmutation.
We already know, from the criteria of the general theory of dimensional
transmutation~\cite{cam:00a}, that
Eq.~(\ref{eq:ISP_eigenvalues3}) implies the existence of a ground state
alone.
However, we will now illustrate the general arguments for this 
particular
problem.

The value of the critical coupling
depends on the angular momentum quantum number $l$
but is independent
of the radial quantum number $n_{r}$.
This, combined with the form of
${\mathcal G}_{n_{r} \, l}(\epsilon)$ in Eq.~(\ref{eq:ISP_G_function}),
imposes very stringent conditions on the existence of bound
states, as we shall see next.

The dependence of $\lambda_{l}^{(\ast)}$ with respect to $l$
requires that {\em only\/}  $l=0$ states (if any)
be allowed, as implied by the following argument.
First, only a finite number of states can exist with
different angular momentum
numbers $l$. In effect,
let us assume the existence
of a given state with angular momentum $l_{0} $;
as the coupling is critically strong, then
$\lambda= \left(l_{0} +\nu_{0} \right)^{2}$,
 so that the potential is ``weak'' and has no bound states
for all $l>l_{0}$---for weak coupling, the unregularized theory
suffices.
In other words, the only allowed states are those with
$0 \leq l \leq l_{0}$.
Let us now see that $\l_{0} \neq 0$ would lead to a contradiction;
in effect, if  $\l_{0} > 0$, then
a state with  $l<l_{0}$ would potentially exist as it  would
correspond to the strong regime; however, as $l \neq l_{0}$,
the coupling  would not be critical with respect to $l$,
rendering the theory ill-defined.
In other words, if the ground state exists,
it should have $l=0$ (as expected) and all
states with  $l>0$ are forbidden.

Therefore,
the ground state is characterized by the quantum numbers
\begin{equation}
( {\rm gs} ) \equiv \left( n_{r}=1, l=0  \right)
\;  ,
\end{equation}
and only hypothetical states with $l=0$
survive the renormalization process as bound states.

The next question is whether states with $l=0$
but $n_{r} \neq 0$ can survive renormalization.
Of course, the ratio of
Eqs.~(\ref{eq:ISP_GS}) and (\ref{eq:ISP_DR_regularized_energies2})
provides the answer:
starting with the ground-state energy,
any other such state would have an energy
overwhelmingly suppressed by an exponential factor,
\begin{equation}
\left|
\frac{E_{n_{r} 0 }  }{E_{_{\rm (gs)}} }
\right|
=
\exp \left[
-2^{4/3}
\,
 \left( C_{n_{r}} - C_{1} \right)
\,
\left(
\lambda^{(\ast)}_{_{\rm (gs)}}
 \right)^{-1/3}
\,
\epsilon^{-1/3}
\right]
 \stackrel{(\epsilon \rightarrow 0, n_{r} >1 )}{\longrightarrow}
0
\;  .
\label{eq:DR_excited_states}
\end{equation}
Furthermore,
for these states, the wave function~(\ref{eq:ISP_w_ext3})
has an ill-defined limit,
so that they cease to exist when $\epsilon \rightarrow 0$.
In short, the singular nature of the potential is responsible
for the destruction of all candidates for a renormalized
bound state, except for the limit of
the ground state of the regularized theory---which,
upon renormalization, becomes the ground state of the system
and acquires the
finite energy value~(\ref{eq:ISP_GS}).

Finally, the ground-state wave function can also be explicitly
derived.
From Eqs.~(\ref{eq:ISP_w_ext3}),
 (\ref{eq:ISP_GS}),
and (\ref{eq:effective_radial_wf}),
it follows that the ground state wave function is given by
\begin{equation}
\Psi_{_{\rm (gs)}}
 ({\bf r})
=
\sqrt{
\Gamma (\nu +1) \,
\left(
\frac{\mu^{2}}{\pi}
\right)^{\nu + 1}
}
\;
\frac{K_{0} (\mu r)}
{ \left( \mu r \right)^{\nu} }
\;  ,
\label{eq:ISP_wf_normalized_renormalized}
\end{equation}
which is similar to that of
the two-dimensional delta-function potential,
with which it coincides when $D_{0}=2$.

\subsection{Scattering Sector for the Inverse Square Potential}
\label{sec:ISP_scatt}

For the scattering sector
of the theory we will match the interior and exterior solutions by means
of the parameter
\begin{equation}
\omega (\Theta_{l})=
\frac{\epsilon}{2}
\,
{\mathcal L} (\Theta_{l})
=
\Theta_{l}
\,
\frac{d \ln {\mathcal C}_{\epsilon/4}(\Theta_{l})}{d \Theta_{l}}
\;  .
\label{eq:omega_def}
\end{equation}
In particular, the exterior scattering problem
is described by the second line in
Eqs.~(\ref{eq:ISP_w_ext2}) and (\ref{eq:ISP_w_ext3}), so that
Eq.~(\ref{eq:log_derivative_ext})
can now be rewritten as
\begin{equation}
\omega^{(>)}
(\Theta_{l})=
\Theta_{l}
\,
\frac{d \ln {\mathcal C}_{\epsilon/4}(\Theta_{l})}{d \Theta_{l}}
\;  ,
\label{eq:omega_plus}
\end{equation}
which is explicitly given by
\begin{equation}
\omega^{(>)}
 (\Theta_{l})
=
\frac{
\widetilde{A}_{l}^{(+)}
\, \Theta_{l} \, H_{\epsilon/4}^{(1) \, \prime} (\Theta_{l})
+
\widetilde{A}_{l}^{(-)} \, \Theta_{l}
\, H_{\epsilon/4}^{(2) \, \prime} (\Theta_{l})
}
{
\widetilde{A}_{l}^{(+)}  \, H_{\epsilon/4}^{(1) } 
(\Theta_{l})
+
\widetilde{A}_{l}^{(-)}  \, H_{\epsilon/4}^{(2) } 
(\Theta_{l})
}
\;  .
\label{eq:omega_ext}
\end{equation}
Equation~(\ref{eq:omega_ext})
can be evaluated from the small-argument behavior of the Hankel
functions~\cite{abr:72},
\begin{eqnarray}
  H^{(1,2)}_{p} (z)
&  \stackrel{(z \rightarrow 0)}{\sim} &
 \pm
\frac{ e^{\mp ip\pi/2} }{\pi i}
\,
\left[
e^{\mp i p \pi/2}
\,
\Gamma (-p)
\left( \frac{z}{2} \right)^{p}
 +
e^{\pm i p \pi/2}
\,
\Gamma (p)
\left( \frac{z}{2} \right)^{-p}
\right]
\,
\left[ 1 + O(z^{2}) \right]
\;  ,
\label{mod_Hankel_small_arg}
\end{eqnarray}
so that
\begin{equation}
\omega_{l}
=
\omega^{(>)}
 (\Theta_{l})
=
\frac{2i}{\pi}
\,
\frac{ \widetilde{A}_{l}^{(+)}
- \widetilde{A}_{l}^{(-)}  }
{ (1+iQ_{l}) \,   \widetilde{A}_{l}^{(+)} + (1-iQ_{l}) \,
\widetilde{A}_{l}^{(-)}  }
\;  ,
\label{eq:omega_explicit}
\end{equation}
where
\begin{equation}
Q_{l}=
\frac{2}{\pi} \left( -c + \ln \Theta_{l} \right)
\;  ,
\label{eq:ISP_Q}
\end{equation}
with $c$ defined in Eq.~(\ref{eq:c_log_derivative}).
Then the scattering matrix
can be derived from the asymptotic expansion of the Hankel functions
as $r \rightarrow \infty$,
which according to  Eq.~(\ref{eq:ISP_w_ext3})
is of the form
\begin{equation}
u(r)
 \stackrel{(r \rightarrow \infty , \epsilon \rightarrow 0) }{\sim}
\sqrt{r} \,
\left[
A_{l}^{(+)}
 H^{(1)}_{l+\nu_{0}}
\mbox{\boldmath\Large  $\left(  \right.$ } \! \! \!
k r + \left( l+\nu_{0} \right) \frac{\pi}{2}
\mbox{\boldmath\Large  $\left.  \right)$ } \! \!
+
A_{l}^{(-)}
 H^{(2)}_{l+\nu_{0}} (  k r )
\mbox{\boldmath\Large  $\left(  \right.$ } \! \! \!
k r + \left( l+\nu_{0} \right) \frac{\pi}{2}
\mbox{\boldmath\Large  $\left.  \right)$ } \! \!
\right]
\;  ,
\label{eq:ISP_scatt_asymptotic_wf}
\end{equation}
where the coefficients
for the $s$ wave ($l=0$) are related via
\begin{equation}
A_{0}^{(\pm)}
=
\widetilde{A}_{0}^{(\pm)}
\,
 e^{\pm i(D_{0}-2)\pi/4}
\;  ;
\label{eq:ISP_A_from_A_tilde}
\end{equation}
then
the $s$-wave scattering matrix elements---from
Eqs.~(\ref{eq:omega_explicit}),
(\ref{eq:ISP_A_from_A_tilde}),
 and (\ref{eq:scatt_matrix_from_coeff})---are given by
\begin{equation}
S_{0}^{(D_{0})} (k)
=  e^{i\nu_{0} \pi}
\,
\widetilde{S}_{0}^{(D_{0})} (k)
\;  ,
\label{eq:ISP_scatt_matrix_from_S_tilde}
\end{equation}
with $\widetilde{S}_{0}^{(D_{0})} (k)
 = \widetilde{A}_{0}^{(+)}/\widetilde{A}_{0}^{(-)}$
of the form
\begin{equation}
 \widetilde{S}_{0}^{(D_{0})} (k)
= -
\frac{1 + i \left(2 \omega_{_{\rm (gs)}}^{-1}/\pi - Q_{_{\rm (gs)}}
\right) } {1 - i \left(2 \omega_{_{\rm (gs)}}^{-1}/\pi -
          Q_{_{\rm (gs)}}
   \right) }
\; ,
\label{eq:ISP_scatt_matrix_tilde}
\end{equation}
and where $\omega_{_{\rm (gs)}}
= \omega_{0}$ and $ Q_{_{\rm (gs)}} = Q_{0}$.

In order to obtain an explicit expression for
the scattering matrix in terms of the energy,
we need to evaluate the coefficient $\omega_{_{\rm (gs)}} $ in
Eq.~(\ref{eq:ISP_scatt_matrix_tilde}).
This program can be implemented  through
the expansion of the dual energy
$\widetilde{\eta}$, defined in Eq.~(\ref{eq:dual_energy_ISP}),
which  should now be derived
again for the scattering sector of the theory.
With that purpose in mind,
making use of Eqs.~(\ref{eq:c_eps}),
(\ref{eq:ISP_p_l_epsilon2}),
and
\begin{equation}
\left| \frac{E}{\mu^{2}} \right|^{-\epsilon/2}
=
1 -
\frac{\epsilon}{2} \ln \left| \frac{E}{\mu^{2}}  \right|
+
O
\left( \epsilon^{2}
\right)
\;  ,
\label{eq:eta_exp}
\end{equation}
 the required expression
becomes (for $l=0$)
\begin{equation}
\widetilde{\eta}
=
p^{2}
\,
 \frac{\lambda}{ \lambda_{_{\rm (gs)}}^{(\ast)}} 
\,
\left\{
 1 +
\left( \lambda_{_{\rm (gs)}}^{(\ast)} \right)^{1/2}
\left[
 \ln \pi + \gamma
+ 2  \left( \lambda_{_{\rm (gs)}}^{(\ast)} \right)^{-1/2}
- \ln  \left| \frac{E}{\mu^{2}}   \right|
\right]
\,  p^{-1}
+ O \left( p^{-2}  \right)
\right\}
\;  .
\label{eq:eta_tilde_scatt}
\end{equation}
In addition, we have already renormalized
 the bound-state
sector of the theory, with the result that the coupling constant
should behave as dictated by Eq.~(\ref{ISP_coupling_reg}),
whence
\begin{equation}
\widetilde{\eta}^{1/2}
=
p
\,
\left\{
 1 +
C_{1}   p^{-2/3}
+
\frac{\epsilon}{4}
\left[
-2c
+ g^{(0)}
- \ln \left| \frac{E}{\mu^{2}}  \right|
\right]
+ o(\epsilon)
\right\}
\;  ,
\label{eq:eta_tilde_scatt2}
\end{equation}
from which it follows that
the variable $
y
=
\widetilde{\eta}^{1/2} \, z_{2}
$, defined by Eq.~(\ref{eq:Debye_variable}),
is of the form~(\ref{eq:order_of_y}),
with
\begin{equation}
x=
C_{1} +
\left[
\left(
-c + \ln \Theta_{_{\rm (gs)}}
\right)
- \frac{1}{2} \ln \left| \frac{E}{ E_{_{\rm (gs)}}}
  \right|
\right]
\,
\left( \lambda_{_{\rm (gs)}}^{(\ast)} \right)^{1/2}
\,
p^{-1/3}
\;  ,
\label{eq:x_scatt}
\end{equation}
and $c$ defined in Eq.~(\ref{eq:c_log_derivative}).
Then the logarithmic derivative~(\ref{eq:log_derivative_int})
can be derived for the scattering sector using an argument
similar to the one employed in Eq.~(\ref{eq:BS_log_derivative_int2}),
i.e., with $\delta= x-C_{1}$,
\begin{eqnarray}
{\mathcal L}^{(<)} (\Theta_{l})
& =  &
- 2^{1/3} \,  p^{2/3}
\,
\frac{     Ai^{\, \prime} \left( - 2^{1/3} x \right) }
{ Ai \left( - 2^{1/3} x \right) }
\,
\left[
1 + O
\left(
p^{-2/3} \right) \right]
\nonumber \\
& = &
\frac{  p^{2/3}  }{ x-C_{1} }
\,
\left[
1 + O \left( p^{-2/3} \right) \right]
\;  .
\label{eq:scatt_log_derivative_int}
\end{eqnarray}
As a consequence,
we conclude that the parameter defined in 
Eq.~(\ref{eq:omega_def})
becomes
\begin{equation}
\omega_{_{\rm (gs)}}
=
\omega^{(<)}
 (\Theta_{_{\rm (gs)}})=
\left[
\left(
-c + \ln \Theta_{_{\rm (gs)}}
\right)
- \frac{1}{2} \ln \left| \frac{E}{ E_{_{\rm (gs)}}}  \right|
\right]^{-1}
\;  .
\label{eq:omega_int_expansion}
\end{equation}

Finally, from Eqs.~(\ref{eq:ISP_Q}),
(\ref{eq:ISP_scatt_matrix_from_S_tilde}),
(\ref{eq:ISP_scatt_matrix_tilde}),
and (\ref{eq:omega_int_expansion}),
the S-matrix reads
\begin{equation}
S_{0}^{(D_{0})} (k)
= e^{i\nu_{0} \pi} \;
\frac{
\ln \left(k^{2}/|E_{_{\rm (gs)}}| \right) + i \pi
}
{
\ln \left(k^{2}/|E_{_{\rm (gs)}}| \right) - i \pi
}
\;  ,
\label{eq:ISP_scatt_matrix2}
\end{equation}
while the phase shifts
are implicitly given by the expression
\begin{equation}
\tan
\mbox{\boldmath\Large  $\left(  \right.$ } \! \! \!
\delta_{0}^{(D_{0})} (k)
- \frac{\pi \nu_{0}}{2}
\mbox{\boldmath\Large  $\left.  \right)$ } \! \!
=
\frac{\pi}{
\ln \left( k^{2}/|E_{_{\rm (gs)}}| \right) }
\;  ,
\label{eq:ISP_phase_shift}
\end{equation}
and  the
partial scattering amplitude reads
\begin{equation}
 a^{(D_{0})}_{0}(k)
=
\frac{1}{2 i k^{(D_{0}-1)/2}}
\,
\frac{
\ln \left( k^{2}/|E_{_{\rm (gs)}}| \right)
\left( e^{i\nu_{0} \pi} - 1 \right)
+
i \pi
\,
\left( e^{i\nu_{0} \pi} + 1  \right)
}
{
\ln \left( k^{2}/|E_{_{\rm (gs)}}| \right)
- i \pi
}
\;  .
\label{eq:ISP_partial_scatt_amplitude}
\end{equation}

Equations~(\ref{eq:ISP_scatt_matrix2}),
(\ref{eq:ISP_phase_shift}), and  (\ref{eq:ISP_partial_scatt_amplitude})
are remarkably similar to
Eqs.~(\ref{eq:2D_delta_scatt_matrix}),
(\ref{eq:2D_delta_phase_shift}), and 
(\ref{eq:delta_scattering_amplitude})
for the two-dimensional delta function potential.
Table~\ref{tab:ISP_scatt}
summarizes the main results of the scattering for an arbitrary
number of dimensions $D_{0}$.

\newtheorem{tab}{\rm TABLE}
\renewcommand{\thetab}{\Roman{tab}}

\begin{center}
\begin{tab}
\hspace{-.05in}. \hspace{.1in}
{\rm Phase Shift $\delta_{0}^{(D_{0})} (k)$,
Scattering Matrix $S_{0}^{(D_{0})} (k)$,
and Partial Scattering Amplitude $  a^{(D_{0})}_{0}(k) $
 for the Inverse Square Potential,
as a Function of the Geometric Dimension $D_{0}$ of
Position Space.}
\label{tab:ISP_scatt}
\end{tab}
\end{center}

\begin{center}
\begin{tabular}{cccc}  \hline\hline
 $D_{0}$ \hspace{.2in} &   $\tan \delta_{0}^{(D_{0})} (k)$ \hspace{.2in}
& $S_{0}^{(D_{0})} (k)$ \hspace{.2in} &
$a^{(D_{0})}_{0}(k)
\,  k^{(D_{0}-1)/2} $ \hspace{.2in}
 \\  \hline
&
&
&
\\
$1 \! \pmod{4}$
\hspace{.2in}
&
$ {\displaystyle
\frac{ \pi-L}{ \pi+L}}$
\hspace{.2in}
&
$ {\displaystyle
\frac{ \left(1+i \right) \pi  +  \left(1-i \right) L }{
\left(1-i \right) \pi  +  \left(1+i \right) L }  }  $
\hspace{.2in}
&
$ {\displaystyle
\frac{  \pi - L }{
\left(1-i \right) \pi  +  \left(1+i \right) L }  }  $
\hspace{.2in}
\\
&
&
&
\\
$2 \! \pmod{4}$
\hspace{.2in}
& ${\displaystyle \frac{\pi}{L}}$
\hspace{.2in}
&
${\displaystyle
\frac{  L +i \pi}{
 L - i \pi }} $
\hspace{.2in}
&
${\displaystyle
\frac{ \pi }{
L  - i \pi }} $
\hspace{.2in}
\\
&
&
&
\\
$3 \! \pmod{4}$
\hspace{.2in}
&
$ {\displaystyle
\frac{L+ \pi}{ L-\pi}}$
\hspace{.2in}
&
${\displaystyle
\frac{ \left(-1+i \right) \pi  +  \left(1+i \right) L }{
-\left(1+i \right) \pi  +  \left(1-i \right) L }  } $
\hspace{.2in}
&
${\displaystyle
\frac{ L + \pi }{
-\left(1+i \right) \pi  +  \left(1-i \right) L }  } $
\hspace{.2in}
\\
&
&
&
\\
$4 \! \pmod{4}$
\hspace{.2in}
& ${\displaystyle -\frac{L}{\pi} }$
\hspace{.2in}
&
${\displaystyle
\frac{ \pi  -i L }{
\pi  + i L }} $
\hspace{.2in}
&
${\displaystyle
-\frac{ L }{
\pi  + i L }} $
\hspace{.2in}
\\
&
&
&
\\
\hline\hline
\end{tabular}
\end{center}

{\small
{\em Note\/}. The
shorthand $L= \ln \left( k^{2}/|E_{_{\rm (gs)}}| \right) $  
is used in this table.}

\bigskip

The above analysis refers to $l=0$.
For all other values $l>0$,
the coupling will be weak, so that the phase shifts
will be given by the values of Eq.~(\ref{ISP_weak_phase_shifts}),
with the condition that $ \lambda =
\lambda^{(\ast)}_{_{\rm (gs)}}  $; then,
\begin{equation}
\left.
 \delta_{l}^{(D_{0})}
\right|_{l \neq 0}
=
 \left[ \left( l+ \nu_{0}  \right)
- \sqrt{ l \left(l +  2 \nu_{0} \right)} \;
\right]  \, \frac{\pi}{2}
\; ,
\label{ISP_renormalized_phase_shifts_l_neq_0}
\end{equation}
and
\begin{equation}
\left.
 S_{l}^{(D_{0})} (k)
\right|_{l \neq 0}
=
\left( -1 \right)^{l}
\,
\exp \left(
i \nu_{0} \, \pi
\right)
\,
\exp \left[
-i \, \pi \, \sqrt{l  \left( l+ 2 \nu_{0} \right) }  \;
\right]
\; ,
\label{ISP_scatt_matrix_l_neq_0}
\end{equation}
which are scale-invariant expressions.

A number of consequences follow from
Eqs.~(\ref{eq:ISP_scatt_matrix2}) and
(\ref{eq:ISP_phase_shift}):
\begin{enumerate}
\item
 The unique pole of the scattering
matrix~(\ref{eq:ISP_scatt_matrix2})
corresponds to the unique bound state.
\item
The phase shifts can be renormalized using a
floating renormalization scale $\mu$, as an alternative
to ground-state renormalization.
Then the $s$-wave phase shift reads simply
\begin{equation}
\frac{1}{\tan
\mbox{\boldmath\Large  $\left(  \right.$ } \! \! \!
\delta_{0}^{(D_{0})} (k) - \pi \nu_{0}/2
\mbox{\boldmath\Large  $\left.  \right)$ } \! \!
}
=
\frac{1}{\tan
\mbox{\boldmath\Large  $\left(  \right.$ } \! \! \!
\delta_{0}^{(D_{0})} (\mu) - \pi \nu_{0}/2
\mbox{\boldmath\Large  $\left.  \right)$ } \! \!
}
+ \frac{1}{\pi}
\ln  \left( \frac{k}{\mu} \right)^{2}
\; .
\label{ISP_scatt_floating_renormalization}
\end{equation}
\item
Equations~(\ref{eq:ISP_partial_scatt_amplitude}),
(\ref{ISP_renormalized_phase_shifts_l_neq_0}),
(\ref{eq:scatt_amplitude}), and
(\ref{eq:partial_wave_amplitude})
give the differential scattering cross section
\begin{equation}
f^{(D_{0})}_{k}(\cos \theta)
=
\frac{ N_{\nu_{0}} }
{2i k^{(D_{0}-1)/2} }
\,
\frac{
\ln \left( k^{2}/|E_{_{\rm (gs)}}| \right)
\left( e^{i\nu_{0} \pi} - 1 \right)
+
i \pi
\,
\left( e^{i\nu_{0} \pi} + 1  \right)
}
{
\ln \left( k^{2}/|E_{_{\rm (gs)}}| \right)
- i \pi
}
+
 \stackrel{ \bigtriangledown  }{f}^{(D_{0})}_{k}
(\cos \theta)
\;  ,
\label{eq:ISP_scatt_amplitude}
\end{equation}
where
\begin{eqnarray}
 \stackrel{ \bigtriangledown  }{f}^{(D_{0})}_{k}
(\cos \theta)
=
\frac{ N_{\nu_{0}} }
{2i k^{(D_{0}-1)/2} }
\,
\sum_{l=1}^{\infty}
 &  & \! \! \! \! \! \!
 \left(\frac{l}{\nu_{0}}+ 1 \right) \,
\nonumber \\
& \times &
\! \!
\left\{
\left( -1 \right)^{l}
\,
e^{
i \nu_{0} \, \pi
}
\,
e^{-i \, \pi \, \sqrt{l  \left( l+ 2 \nu_{0} \right) } }
- 1
\right\}
\,
C_{l}^{(\nu_{0})} ( \cos \theta)
\;  .
\label{eq:ISP_scatt_amplitude_l_neq_0}
\end{eqnarray}
\item
All the relevant quantities are logarithmic with respect to
the energy and agree with the predictions of
generalized dimensional analysis~\cite{cam:00a}.
\item
Equations~(\ref{eq:ISP_scatt_matrix2}),
(\ref{eq:ISP_phase_shift}),
and (\ref{eq:ISP_partial_scatt_amplitude})
relate the bound-state and scattering sectors of the theory and show
that the inverse square potential is renormalizable.
\end{enumerate}

\section{CONCLUSIONS}
\label{sec:conclusions}

In this paper, we have uncovered a number of remarkable analogies
between the two-dimensional delta-function and inverse square 
potentials.
In addition to displaying their characteristic
 transmuting behavior,
with all the ensuing implications,
we have explicitly seen that they share the following
properties:

(i)
Unusual
boundary conditions at the origin.

(ii) Characteristic
critical couplings (coincident for $D_{0}=2$)
that determine the possible regimes of each potential.

(iii)
Only one bound state in the renormalized theory,
even though this is achieved through different mechanisms
(the delta-function potential always generates a unique state,
while the inverse square potential annihilates
all the regularized excited states by exponential suppression).

(iv) Almost identical ground-state wave functions (up to
the normalization constant).

(v) Similar $s$-wave
scattering matrix elements---they are proportional,
differing only upon an extra $D_{0}$-dependent phase factor
for the inverse square potential.

(vi) Characteristic logarithmic behavior
of  $s$-wave scattering quantities.

It should be noticed that the bound-state sectors look
essentially identical in both theories, while the $s$-wave
scattering sectors are identical for $D=2$ and almost identical
for $D \neq 2$.

However, there a number of differences as well.
Due to its zero-range nature, the two-dimensional delta function
only scatters $s$ waves, while the inverse square potential,
due to its infinite range,
scatters all other angular-momentum channels in
a scale-invariant (energy-independent) way.

Moreover, these results are independent
of the regularization technique; in particular,
they are in perfect agreement
with the $D_{0}$-dimensional generalization~\cite{cam:00f}
of the cutoff-renormalization method
of Ref.~\cite{gup:93}.

Finally, the techniques used in this paper
could be easily generalized.
For example, one could consider the generalized
inverse square potential
\begin{equation}
V({\bf r})
= - \lambda
\,
\frac{
v
 ( \Omega^{({D})}  )}{ r^{2} }
\;  ,
\label{eq:gralized_ISP}
\end{equation}
with a dimensionless
function $v (\Omega^{({D})} )$ that
depends on the $D$-dimensional solid angle $ \Omega^{({D})}$;
the angular part of the solution
could be properly modified, but the
basic scaling relationships would remain the same.
In principle, this strategy could be conveniently used for the dipole
potential~\cite{lev:67,des:94} and for other forms of angular dependence
in Eq.~(\ref{eq:gralized_ISP}).

\appendix

\section{ROTATIONAL INVARIANCE: CENTRAL POTENTIALS IN $D$ DIMENSIONS}
\label{sec:central}

In this appendix
we will enforce the condition of rotational
invariance and use the notation and definitions of hyperspherical 
coordinates
as introduced in Ref.~\cite{cam:00a}.

As usual, a central potential $V(r)$ is defined to be
rotationally invariant; thus, its functional form is
 independent of the angular variables
in $D$-dimensional hyperspherical coordinates.
Its associated symmetry group is  $SO(D)$, with a quantum mechanical
representation generated by the $D(D-1)/2$ generalized
angular momentum operators
$x_{i}  p_{k}-x_{k}p_{i} $ (with $i<k$).
The quadratic Casimir operators
\begin{eqnarray}
L_{j}^{2}
& =  &
\sum_{k>i=j}^{D-1}
\left(
x_{i}  p_{k}-x_{k}p_{i}
\right)^{2}
\nonumber \\
& =  &
- \hbar^{2}
\sum_{k=j}^{D-1}
\left(
\prod_{i=j}^{k-1} \sin^{2} \theta_{i}  \right)^{-1}
\left[
\frac{\partial^{2} }{\partial \theta_{k}^{2} }
+ (D-k-1) \cot \theta_{k}
\frac{\partial }{\partial \theta_{k} }
\right]
\;
\label{eq:gralized_ang_momentum}
\end{eqnarray}
(with $j=1, \ldots D-1$)
commute with all the generators of the Lie algebra
and with all possible rotations.
Each $L_{j}^{2}$ represents
the ``total'' $D$-dimensional angular momentum
squared  in the subspace spanned by the Cartesian coordinates
$(x_{j}, \ldots , x_{D})$, and
is characterized by the properties that
it commutes with the Hamiltonian of any central potential
and that it satisfies the eigenvalue equation
\begin{equation}
L_{j}^{2}   \,
| E , L >
=
l_{j} (l_{j} +D-j-1) \hbar^{2}
\,
| E , L >
\;  ,
\label{eq:gralized_ang_momentum_eigenvalues}
\end{equation}
where the eigenstates $| E , L > $ are
labeled by the energy and by the
 collective index of generalized angular
momentum quantum numbers
\begin{equation}
L = \left( l\equiv l_{1}, l_{2}, \ldots, l_{{D}-2},
m \equiv l_{{D}-1} \right)
\;  .
\label{eq:angular_momentum_q_numbers_def}
\end{equation}
The eigenstates  $| E , L > $ lead to
the generalization of the usual 3D factorization of the position wave
function, which  can be written
as~\cite{lou:60}
\begin{equation}
\Psi_{E L}
(r, \Omega^{(D)})
=
<r, \Omega^{(D)}
| E , L > =
R_{El} (r)
\,
Y_{L} (\Omega^{(D)})
\; ,
\label{eq:separation_Psi}
\end{equation}
in terms of the  hyperspherical harmonics
$Y_{L} (\Omega^{({D})})= <\Omega^{(D)}
| L > $.
Notice that
the peculiar number  $m$ is associated
with rotations on the $(x_{D-1},x_{D})$ plane,
which are characterized by the azimuthal angle $\phi \equiv 
\theta_{D-1}$;
the corresponding operator is usually chosen to be
\begin{equation}
L_{D-1}=
x_{D-1}\,  p_{D} -
x_{D} \, p_{D-1}
=
\frac{\hbar}{i} \frac{\partial}{\partial  \phi}
\;  ,
\label{eq:L_D-1}
\end{equation}
instead of $L_{D-1}^{2}$,
with eigenvalues $m \hbar$ (both positive and negative).
In addition,
the generalized angular momentum quantum numbers satisfy
 the constraints
\begin{equation}
0 \leq |m| \leq l_{D-2} \leq l_{D-3} \leq \ldots \leq
l_{3} \leq l_{2} \leq l
\;  .
\label{eq:angular_momentum_q_numbers_constraints}
\end{equation}

Moreover, the angular part of the Laplacian~\cite{cam:00d},
\begin{equation}
\Delta_{\Omega^{({D})}}
=
\sum^{{D}-1}_{j=1}
\left[
\left( \prod_{k=1}^{j-1} \sin^{2} \theta_{k} \right)
\, \sin^{{D}-j-1} \theta_{j}
\right]^{-1}
\frac{\partial}
{\partial\theta_{j}}
\left(
\sin^{{D}-j-1} \theta_{j}  \,
\frac{\partial}{\partial\theta_{j}}
\right)
\; ,
\label{eq:angular_Laplacian}
\end{equation}
from Eq.~(\ref{eq:gralized_ang_momentum}),
can be simply written in terms of
the total angular momentum $L^{2}=L_{1}^{2}$, namely,
\begin{equation}
\Delta_{\Omega_{{D}}}
 = -\frac{L^{2}}{\hbar^{2}}
\; ,
\end{equation}
so that Eq.~(\ref{eq:gralized_ang_momentum_eigenvalues})
leads to
\begin{equation}
\Delta_{\Omega^{({D})}}
\,
Y_{L} (\Omega^{({D})})
= -
l (l + D - 2) \,
Y_{L} (\Omega^{({D})})
\label{eq:angular_Laplacian_eigenvalue_eq}
\; .
\end{equation}
In particular, the solutions of the angular
part of Laplace's equation that are
independent of all angles but $\theta_{1}$
are the hyperspherical harmonics with
 $l_{2}= \ldots =l_{D-2} =m=0$, for which
Eq.~(\ref{eq:angular_Laplacian_eigenvalue_eq}) reduces to the
ultraspherical differential equation, with regular solutions given by
the Gegenbauer (or ultraspherical) polynomials
$C_{l}^{(\nu)}(\cos \theta_{1})$,
defined in terms of its generating function
(see Refs.~\cite{abr:72,som:64})
\begin{equation}
\left(1-2tz +z^{2} \right)^{-\nu} =
\sum_{l=0}^{\infty} C_{l}^{(\nu)} (t) \, z^{l}
\;  .
\label{eq:Gegenbauer_GF}
\end{equation}
This is all the background needed for our work;
the reader may consult Ref.~\cite{lou:60}
for general proofs and  Refs.~\cite{erd:62,cas:65}
for explicit expressions of the  hyperspherical
harmonics.

Equation~(\ref{eq:angular_Laplacian_eigenvalue_eq})
justifies a posteriori the use of the notation $R_{El}(r) $
introduced in Eq.~(\ref{eq:separation_Psi}), which assumes that
$R (r) $ depends only on the angular momentum  quantum number
$l$ (but not on $l_{2}, \ldots , l_{D-2}, m$),
in addition to the energy $E$.
In effect, after isolating the angular factor that is common
to all central potentials, the wave function $R_{El}(r) $
satisfies the equation
\begin{equation}
\left[
\Delta_{r}^{({D})} -
\frac{l(l+ D -
2)}{r^{2}}+V(r)
\right] R_{El}(r)
= E
R_{El}(r)
\; ,
\label{eq:central_eff_Schr}
\end{equation}
with a radial Laplacian
\begin{eqnarray}
\Delta^{({D})}_{r}
& = & \frac{1}{ r^{{D}-1} }
\frac{\partial}{\partial r}
\left( r^{{D}-1}
\frac{\partial}{\partial r} \right)
\label{eq:radial_Laplacian}
\\
& = &
\frac{1}{r^{({D} -1)/2} }
\frac{d^{2}}{dr^{2}}
\left[ r^{({D} - 1)/2} \right]
+\frac{(D - 1)(D - 3) }{4r^{2}}
\; .
\end{eqnarray}
As a result, the function
\begin{equation}
u_{El}(r)
= R_{El}(r) \,
r^{  (D -1) /2}
\label{eq:effective_radial_wf}
\end{equation}
satisfies a one-dimensional Schr\"{o}dinger equation
\begin{equation}
\left[
-\frac{d^{2}}{dr^{2}}
+ V(r) + \frac{\Lambda_{l, {D} }}{r^{2}}
\right]
u_{El}(r)
= E \,
u_{El}(r)
\;  ,
\label{eq:radial_Schr}
\end{equation}
in which the centrifugal  barrier
is characterized by the
effective coupling constant
\begin{eqnarray}
\Lambda_{l, {D}}
& = &
l(l+ D - 2)+
(D - 1)(D - 3)/4
 \\
& = &
 (l+\nu)^{2} - 1/4
\;  ,
\label{eq:cf_coupling}
\end{eqnarray}
where
$\nu$ is the variable defined in Eq.~(\ref{eq:nu_def}).
In our work, Eq.~(\ref{eq:cf_coupling}) is most useful,
particularly because, for the potentials of interest
in this work, it provides a straightforward connection with a family
of Bessel differential equations.

 Equation~(\ref{eq:cf_coupling}) depends on the number of
dimensions of the space only through the combination $l+ \nu$; this
amounts to the remarkable phenomenon
known as interdimensional dependence~\cite{van:73}:
the solutions for any two problems
related via
$l+ \nu = l^{\prime} + \nu^{\prime}$, with $\nu \neq \nu^{\prime}$
(i.e., $D \neq D^{\prime}$)
 are identical.

A remark about notation is in order.
In most contexts, it is customary
to drop the subscript $E$ labeling the wave functions in
Eqs.~(\ref{eq:separation_Psi}),
(\ref{eq:central_eff_Schr}),
(\ref{eq:effective_radial_wf}), and
(\ref{eq:radial_Schr}), i.e., to write
$R_{El} (r) \equiv R_{l} (r) $
and
$u_{El} (r) \equiv u_{l} (r) $.
Alternatively, for the bound-state sector,
one often writes
$R_{n_{r}l} (r)$,
 $u_{n_{r}l} (r)$, and  $E_{n_{r}l}$,
where $n_{r}$ is the radial quantum number.

Finally, in order to have a well-defined problem,
 boundary conditions are needed both at infinity and at the origin.
At infinity, the bound-state solutions should go to zero in order to
ensure their integrability, while the scattering solutions are subject 
to the usual requirements~\cite{cam:00e}.
On the other hand, the boundary condition at $r=0$
requires further analysis.

In fact, the boundary condition at the origin is the basis for
the classification of potentials into the regular and singular
families, as discussed in
Appendix~\ref{sec:powerlaw}.
 In this framework,
 regular and semi-regular potentials are characterized by the limit
\begin{equation}
r^{2} V(r)
\stackrel{ r \rightarrow 0}{\longrightarrow}
0
\;  ,
\end{equation}
so that,
 near the origin, both the potential and total energy terms
are negligible
and the limiting form of the wave function
becomes asymptotically a solution of the radial part of
Laplace's equation, i.e.,
\begin{equation}
R_{l}(r) 
 \stackrel{(r \rightarrow 0 )}{\sim}
\frac{u_{l}(r)}{r^{( {D} - 1)/2  }}
=
\mbox{\boldmath\large  $\left\{  \right.$ } \! \! \!
r^{l}
\mbox{\boldmath\large  $,$ }  \!
r^{-(l + 2 \nu)}
\mbox{\boldmath\large  $\left.  \right\}$ } \! \!
\;    ,
\label{eq:Laplace_solutions}
\end{equation}
where
$\mbox{\boldmath\large  $\left\{  \right.$ } \! \! \!
\mbox{\boldmath\large  $,$ } \! \! \!
\mbox{\boldmath\large  $\left.  \right\}$ } \! \!
$
stands for linear combination.
In Eq.~(\ref{eq:Laplace_solutions}),
the first component is acceptable but the second
one should be discarded because
$
\nabla^{2}_{{D}}
 \left[
 Y_{L} (\Omega_{{D}})
/r^{l+ {D} -2 }
\right]$
is proportional to the multipole density of order $l$,
i.e., it involves
derivatives of the delta function
$\delta^{({D})} ({\bf r})$
of order $l$.
Therefore, for regular potentials, there is a criterion
for the selection between the two linearly
independent solutions  of Eq.~(\ref{eq:radial_Schr}),
and this provides the boundary condition
\begin{equation}
R_{l}(r)
\propto
r^{l}
\;  .
\label{eq:asympt_BC}
\end{equation}
In practice, it is sufficient to consider the weaker boundary
condition
\begin{equation}
u_{ l} (0) = 0
\;  .
\label{eq:BC_at_origin}
\end{equation}
As discussed
in Subsection~\ref{sec:ISP_loss_BC}, the source of the unusual properties
 of critical (dimensionally transmuting) and singular potentials
is the ``loss'' of the
boundary condition~(\ref{eq:BC_at_origin}).

\section{ROTATIONAL INVARIANCE: PARTIAL-WAVE ANALYSIS IN $D$ DIMENSIONS}
\label{sec:D-dim_partial_wave_analysis}

For our work, we need another aspect of rotational invariance
in $D$ dimensions: the
expansion in $D$-dimensional partial waves, which
is developed next.
A spherical wave state $|E, L>$ is  represented by a  solution
of the radial Schr\"{o}dinger equation for a free particle;
then Eq.~(\ref{eq:central_eff_Schr}),
when $V(r)=0$,
has solutions proportional to the
Bessel functions $Z_{l +\nu} (kr)$, with $\nu=D/2 -1$,
where
$Z=J$, or $N$, or $H^{(1,2)}$;
explicitly,
\begin{equation}
R_{l}(r) \propto
z_{l,\nu} (x)
=
\sqrt{\frac{\pi}{2}} \,
x^{-\nu}
  Z_{l+\nu}(x)
\;  ,
\end{equation}
where the  ultraspherical Bessel functions
 admit the asymptotic expansions
\begin{equation}
z_{l,\nu} (x)
 \stackrel{(x \rightarrow \infty ) }{\sim}
x^{-(D-1)/2}
\,
\tau \left( x +
\gamma_{D}
- l \frac{\pi}{2} - \frac{\pi}{2}
\right)
\;  ,
\end{equation}
in which $\tau (\xi)$ is the corresponding trigonometric function
(i.e., $\cos \xi$ for $j(\xi)$, $\sin \xi$ for $n(\xi)$,
and $\exp (\pm i \xi)$ for $h^{(1,2)}(\xi)$),
and $\gamma_{D} = (3-D)\pi/4$ (from Ref.~\cite{cam:00a}).
In particular, the regular free-particle solution
of Eq.~(\ref{eq:central_eff_Schr})
(based on its behavior at the origin)
is provided by
$j_{l ,\nu} (kr)$.
On the other hand, the angular part
leads to the Gegenbauer (or ultraspherical) polynomials
$C_{l}^{(\nu)}(\cos \theta)$ defined
through Eq.~(\ref{eq:Gegenbauer_GF}).
In short, the regular solution is of the form
$j_{l,D/2-1}
 (kr)\,  C_{l}^{(D/2-1)} (\cos \theta)$.

The partial-wave expansion for central potentials in $D$ dimensions
then proceeds
in complete analogy to the three-dimensional case~\cite{sak:85}.
This can be accomplished by considering the transition
from a plane-wave state $|{\bf k}>$, represented by the
wave function $e^{ikr \cos \theta}$ (we will
choose $\theta \equiv \theta_{1}$ for the sake of simplicity),
 to a spherical-wave
state $|E, L>$.
The coefficients of the transition can be found from the identity
 (Ref.~\cite{abr:72}, p. 363)
\begin{equation}
e^{i x \cos \theta}
=
\Gamma (\nu)
 \left( \frac{x}{2} \right)^{-\nu}
\sum_{l=0}^{\infty} \left( l+ \nu \right) \,
i^{l}
\,
J_{l+ \nu} (x)
\,
C_{l}^{(\nu)} ( \cos \theta)
\; ,
\end{equation}
with $x=kr$;
this implies that
\begin{equation}
e^{i kr \cos \theta}
=N_{\nu}
\,
\sum_{l=0}^{\infty}
  \left(  \frac{l}{\nu} + 1 \right)
\, i^{l} \,
j_{l,\nu} (kr)
C_{l}^{(\nu)} ( \cos \theta)
\; ,
\label{eq:Rayleigh}
\end{equation}
which is the $D$-dimensional generalization of Rayleigh's formula,
with
\begin{equation}
N_{\nu}
= 2^{\nu}\,  \Gamma (\nu +1) \, \sqrt{\frac{2}{\pi}}
\;  .
\label{eq:N_nu}
\end{equation}
Equation~(\ref{eq:Rayleigh})
straightforwardly reduces to the familiar result for $D=3$
($\nu=1/2$). The case $D=2$ ($\nu=0$) appears to be singular,
but also reproduces the known results~\cite{mor:53e}
when the following replacements are made:
(i) $C_{l}^{(0)} (t )
= \lim_{\nu \rightarrow 0}
C_{l}^{(\nu)} (t )/\nu$
(Ref.~\cite{abr:72});
(ii)  $C_{l}^{(0)} (\cos \theta )
=2 \cos (l\theta)/l$  for $l \neq 0$ (this is proportional
to a Chebyshev polynomial, Ref.~\cite{abr:72})
but $C_{0}^{(0)} (\cos \theta )=1$;
and (iii)
most often,
the sum is extended from $m=-\infty$ to $m=\infty$,
with $m= \pm l$, and the factor of
2 in (ii) is removed (otherwise this factor is kept as the Neumann
number, $\epsilon_{l}= 2$ for $l \neq 0$, but
$\epsilon_{0}= 1$).

As the transition operator $T$ commutes with the generalized
angular momentum operators, it has a diagonal form in the angular
momentum eigenbasis, with elements $T_{l}^{(D)}(k)$; thus, one can
expand its matrix elements $<{\bf k}| T|{\bf k}'>$ in terms
of $T_{l}^{(D)}(k)$, with
\begin{equation}
T_{l}^{(D)}(k)
= -
\frac{ S_{l}^{(D)} (k) - 1} {2 \pi i}
\;  ,
\label{eq:T_matrix_def}
\end{equation}
where the
 scattering matrix elements
\begin{equation}
S_{l}^{(D)} (k)
=
\exp \left[2i \delta_{l}^{(D)} (k)
\right]
=
\frac{
1 + i \tan \delta_{l}^{(D)} (k) }
{1 - i \tan \delta_{l}^{(D)} (k) }
\;  
\label{eq:scatt_matrix_def}
\end{equation}
are usually expressed in terms of the scattering phase shifts
$\delta_{l}^{(D)} (k)  $.
Then the corresponding scattering amplitude $f^{(D)}_{k}(\cos \theta) $
admits a straightforward
expansion
\begin{equation}
f^{(D)}_{k}(\cos \theta) =
N_{\nu}
\,
\sum_{l=0}^{\infty}
  \left(  \frac{l}{\nu} + 1 \right)   \,
a^{(D)}_{l} (k)
\,
C_{l}^{(\nu)} ( \cos \theta)
\;  ,
\label{eq:scatt_amplitude}
\end{equation}
with an $l$th partial-wave amplitude
\begin{equation}
a^{(D)}_{l} (k)
=
-
\frac{\pi}{ k^{(D-1)/2} } \;  T_{l}^{(D)}(k)
=
\frac{ \exp \left[2i \delta_{l}^{(D)} (k) \right]
-1}{2i k^{(D-1)/2} }
\;
\label{eq:partial_wave_amplitude}
\end{equation}
that provides the asymptotic form of the wave function
\begin{eqnarray}
\Psi ({\bf r})
 \stackrel{(r \rightarrow \infty ) }{\sim}
\frac{N_{\nu}}{(kr)^{(D-1)/2}}
\,
\sum_{l=0}^{\infty}
 \,
\,  \left(  \frac{l}{\nu} + 1 \right)
 & \!
 & \! \! \! \! \!
\exp \left\{
i \left[  \delta_{l}^{(D)} (k) + l \pi/2 \right] \right\}
\,
\nonumber \\
& \times &
\sin     \left[ kr+\gamma_{D} - l \pi/2 + \delta_{l}^{(D)} (k) \right]
\,
C_{l}^{(\nu)} ( \cos \theta)
\;  .
\label{eq:asympt_wf}
\end{eqnarray}

As a practical matter, the scattering matrix can be computed directly
from the asymptotic
expansion of the exact solution to the problem,
i.e.,
\begin{equation}
R_{l} (r)
\equiv
R_{El} (r)
 \stackrel{(r \rightarrow \infty ) }{\sim}
A_{l}^{(+)} h_{l,\nu}^{(1)} (kr)
+
A_{l}^{(-)} h_{l,\nu}^{(2)} (kr)
\;  ,
\label{eq:asympt_exact_sol}
\end{equation}
which yields
\begin{equation}
S_{l}^{(D)} (k)=
\frac{ A_{l}^{(+)} } {A_{l}^{(-)} }
\;  .
\label{eq:scatt_matrix_from_coeff}
\end{equation}
Finally, the  total scattering cross section
can be obtained directly
by integration of $|f^{(D)}_{k}(\cos \theta) |^{2}$,
with the result
\begin{equation}
\sigma_{D}(k) =
\frac{2 \Omega_{D-1}}{k^{D-1}}
\sum_{l=0}^{\infty}
 \left( 2l + 2 \nu  \right) \,
\frac{ \Gamma (l+ 2 \nu)}{l!}  \,
\sin^{2} \delta_{l}^{(D)} (k)
\;  .
\label{eq:total_cross_section}
\end{equation}

\section{DUALITY TRANSFORMATION
FOR POWER-LAW POTENTIALS}
\label{sec:powerlaw}

In this appendix,
which we have adapted from
Ref.~\cite{qui:79},
 we will consider the class of central power-law potentials
\begin{equation}
V(r)=
{\rm sgn} \left( \beta \right)
\,  \lambda \, r^{\beta}
\; ,
\label{eq:power_law_pot_def}
\end{equation}
where the sign
is chosen so that $\lambda >0$ corresponds to attractive potentials.
According to their behavior
at the origin, they can be classified into the following categories.
\begin{enumerate}
\item Regular potentials: $\beta \geq 0$.
\item Semi-regular potentials: $-2 < \beta < 0$.
\item Critically singular potential: $\beta=-2$
\item Strictly singular  potentials: $\beta < -2 $.
\end{enumerate}
What physically
characterizes the singular potentials is that the centrifugal
barrier fails to be the dominant term at the origin; mathematically,
 the Hamiltonian loses its self-adjoint
character~\cite{gal:90}.
In particular, the critical potential
is the one that generates  dimensional transmutation.

The central theme of this appendix is the
existence of a remarkable duality transformation ${\mathcal D}$,
which relates the potentials in the regular and semi-regular families;
in particular,
${\mathcal D}$ establishes a one-to-one correspondence
between the exponents $\beta$ in the intervals $(-2,0]$
 and $[0, \infty)$.
As we will see below, if
\begin{equation}
{\mathcal D}
\mbox{\boldmath\Large  $\left(  \right.$ } \! \! \!
{\rm sgn} \left( \beta \right)
\,  \lambda \, r^{\beta}
\mbox{\boldmath\Large  $\left.  \right)$ } \! \!
=
{\rm sgn} \left( \widetilde{\beta} \right)
\, \widetilde{\lambda}
\,  r^{\widetilde{\beta}}
\;  ,
\label{eq:duality_potentials}
\end{equation}
then the corresponding relation between exponents reads
\begin{equation}
(\beta+2) ( \widetilde{\beta} +2 ) = 4
\; .
\label{eq:duality_exp}
\end{equation}
Equations~(\ref{eq:duality_potentials})
and (\ref{eq:duality_exp})
exhibit the following properties:
(i ) ${\mathcal D} $ is idempotent; (ii) $-2 <\beta< 0$ if and only if
$0 <\widetilde{\beta} < \infty$,
so that  ${\rm sgn} \left( \beta \right) =
- {\rm sgn} ( \widetilde{\beta} ) $;
(iii)
the Coulomb potential ($\beta=-1$)
is the dual of  the harmonic oscillator ($\beta=2$);
(iv) its limiting exponents ($\beta=-2$ and $\beta=\infty$)
define the inverse square potential as the dual
of the infinite hyperspherical potential;
and (v) the constant potential or free-particle case is self-dual.

Let us now show that
Eq.~(\ref{eq:duality_exp}) represents the only nontrivial duality
transformation implemented by a scale transformation for
arbitrary power-law potentials.
In general,
under the  transformation
\begin{equation}
\left\{
   \begin{array}{l}
 r =  f(\varrho)
 \nonumber \\
u (r)   =  \widetilde{u} (\varrho) \, g(\varrho)
\end{array}
\right.
\; ,
\label{eq:transf_Schr1}
\end{equation}
from a radial wave function $u(r)$ to $\widetilde{u} (\varrho)$,
 the Schr\"{o}dinger equation
\begin{equation}
\left[ \frac{d^{2}}{dr^{2}}
+ E -  V(r)
- \frac{(l + \nu)^{2}
 - 1/4}{r^{2}}
\right]  u_{l}(r) = 0
\;  
\label{eq:radialSchr_general}
\end{equation}
is mapped into a transformed equation of the same form,
without first-order derivative term, if and only if
\begin{equation}
\left[ g(\varrho) \right]^{2} \propto f^{\prime} (\varrho)
\; ,
\end{equation}
where the prime, as usual, denotes the derivative;
then
\begin{equation}
\left\{
 \frac{d^{2} }{d\varrho^{2}}
+
\left[ f^{\prime} (\varrho) \right]^{2}
 \left[   E -  V
\mbox{\boldmath\large  $\left(  \right.$ } \! \!
 f(\varrho)
 \!  \mbox{\boldmath\large  $ \left.  \right)$ }
 \right]
- {\mathcal L}_{1}^{2}(\varrho) (l + \nu)^{2}
+ {\mathcal L}_{2}^{\prime} (\varrho)
- {\mathcal L}_{2}^{2} (\varrho)
\right\}
 \widetilde{u}
  (\varrho) = 0
\;  ,
\label{eq:radialSchr_general_transformed}
\end{equation}
where
\begin{equation}
{\mathcal L}_{j}
(\varrho)
=
\frac{1}{j!} \,
 \left( \frac{d \ln }{d \varrho} \right)^{j} f (\varrho)
\;
\end{equation}
(for $j=1,2$).
Even though Eqs.~(\ref{eq:transf_Schr1})
and (\ref{eq:radialSchr_general_transformed})
are fairly general and have many applications, the scale
transformation
\begin{equation}
\left\{
   \begin{array}{l}
 r =  \varrho^{\alpha}
\\ \nonumber
u (r)   =  \widetilde{u} (\varrho) \varrho^{(\alpha -1)/2}
\end{array}
\right.
\; ,
\label{eq:transf_Schr2}
\end{equation}
leading to
\begin{equation}
\left\{
 \frac{d^{2} }{d \varrho^{2}}
+
\alpha^{2} \varrho^{2(\alpha-1)}
 \left[E - V( \varrho^{\alpha})  \right]
- \frac{ \alpha^{2} (l + \nu)^{2} - 1/4 }{ \varrho^{2} }
\right\}
 \widetilde{u} (\varrho) = 0
\;  ,
\label{eq:radialSchr_scale_transformed}
\end{equation}
can immediately provide the desired connection
for the  power-law potentials of Eq.~(\ref{eq:power_law_pot_def}).
In effect,
unless the trivial transformation $\alpha=1$ is allowed, the only
way of bringing the middle term in
Eq.~(\ref{eq:radialSchr_scale_transformed})
to the required transformed form, $\widetilde{E}- 
\widetilde{V}(\varrho)$,
is to perform the exchange
\begin{eqnarray}
 -V(r)  & = & - {\rm sgn} \left( \beta \right) \,
\lambda \, r^{\beta}  \rightarrow  \widetilde{E} \\
E & \rightarrow &
- \widetilde{V} (\varrho) =
- {\rm sgn} \left( \widetilde{\beta} \right) \,
\widetilde{\lambda} \varrho^{\widetilde{\beta}}
\; ;
\end{eqnarray}
this ``crossing'' implies that the exponents be constrained by
\begin{equation}
\alpha = - \frac{\widetilde{\beta}}{\beta}
= \frac{2}{\beta +2}
\; ,
\label{eq:alpha}
\end{equation}
which is equivalent to the anticipated
duality relation~(\ref{eq:duality_exp}).
In conclusion,
the required coordinate and  wave function substitutions are
\begin{equation}
\left\{
   \begin{array}{l}
r  =   \varrho^{2/(\beta + 2)}
 \nonumber  \\
u (r)   =  \widetilde{u} (\varrho)\,
 \varrho^{-\beta/2(\beta + 2)}
\end{array}
\right.
\; ,
\label{eq:transf_Schr3}
\end{equation}
which transform
 Eq.~(\ref{eq:radialSchr_general}) into
\begin{equation}
\left[
\frac{d^{2}}{d\varrho^{2}} + \widetilde{E}
-
{\rm sgn} \left( \widetilde{\beta} \right)
\,  \widetilde{\lambda}  \varrho^{ \widetilde{\beta} }
- \frac{
\left(\widetilde{l} + \widetilde{\nu} \right)^{2}
- 1/4}{\varrho^{2}}
\right]  \widetilde{u}_{l}(\varrho) = 0
\;  ,
\label{eq:radialSchr_powerlaw2}
\end{equation}
with a dual energy eigenvalue
\begin{equation}
\widetilde{E} = - {\rm sgn} \left( \beta \right)
\,
\lambda \alpha^{2}
\; ,
\end{equation}
dual coupling
\begin{equation}
\widetilde{\lambda} = {\rm sgn} \left( \beta \right)
\,   E \, \alpha^{2}
\; ,
\end{equation}
and dual angular momentum quantum number
$\widetilde{l}$, such that
\begin{equation}
\widetilde{l}
 + \widetilde{\nu}  =
\alpha ( l + \nu )
\;  ,
\end{equation}
where $\alpha$ is explicitly given by Eq.~(\ref{eq:alpha}), and
we also allow for the possibility of a
change in the number of dimensions,
according to $\widetilde{\nu}= \widetilde{D}/2 -1$.
In particular,
Eq.~(\ref{eq:radialSchr_powerlaw2}) shows that
the dual potential is, simply,
\begin{equation}
\widetilde{V}  (\varrho)
=
- E \, \alpha^{2} \,
\varrho^{ \widetilde{\beta} }
\;  .
\label{eq:dual_pot}
\end{equation}

Finally, the dimensionless form of
 Eq.~(\ref{eq:radialSchr_powerlaw2}) can be obtained
by introducing an inverse length scale
from the original problem through the energy, i.e.,
\begin{equation}
\kappa = \sqrt{|E|}
\;  ;
\end{equation}
then
the resulting transformation
involves the dimensionless variables
\begin{equation}
\left\{
   \begin{array}{l}
z = \kappa^{1/\alpha} \, \varrho
\\ \nonumber
w (z) = \kappa^{-(D+1-1/\alpha)/2} \, \widetilde{u} ( 
\kappa^{-1/\alpha} z)
\end{array}
\right.
\; ,
\end{equation}
in terms of which
\begin{equation}
\left\{
   \begin{array}{l}
\kappa r =  z^{2/(\beta + 2)}
\\ \nonumber
\kappa^{-D/2} u (r)  =  w (z)  \,
z^{-\beta/2(\beta + 2)}
\end{array}
\right.
\;
\label{eq:transf_Schr4}
\end{equation}
and the transformed Schr\"{o}dinger equation becomes
\begin{equation}
\left[ \frac{d^{2}}{dz^{2}}  - {\rm sgn} \left( \beta \right)
 \lambda \, \alpha^{2} \kappa^{-2/\alpha}
+ \sigma \alpha^{2}  z^{ \widetilde{\beta} }
 - \frac{ (\widetilde{l} + \widetilde{\nu} )^{2} - 1/4}{z^{2}}
\right]  w_{l}(z) = 0
\;  ,
\label{eq:radialSchr_powerlaw3}
\end{equation}
where $\sigma= {\rm sgn} (E)$.
Two important remarks immediately follow
from Eq.~(\ref{eq:radialSchr_powerlaw3}):
(i)
the ensuing dimensionless energy equation,
\begin{equation}
\widetilde{\eta}= - {\rm sgn} \left( \beta \right)
 \, \lambda \, \alpha^{2} |E|^{-1/\alpha}
\;  ,
\label{eq:eta_alpha}
\end{equation}
is the basis for the relation between the energy eigenvalue problems
of the corresponding dual potentials
(for example, for $\beta=-1$, $\widetilde{\beta}=2$,
$\alpha=2$,
it provides the well-known connection between the Coulomb
potential and the harmonic oscillator, in spaces of any
number of dimensions);
and (ii)
the dimensionless dual potential
\begin{equation}
\widetilde{\mathcal V} (z) =
-  \sigma \alpha^{2}  z^{ \widetilde{\beta} }
\;
\end{equation}
satisfies the sign relation
${\rm sgn}(\widetilde{\mathcal V})  = - {\rm sgn}(E)$,
which, combined with Eq.~(\ref{eq:eta_alpha}),
leads to a one-to-one correspondence between the bound-state
sectors and between the scattering sectors of the dual potentials.

For our work,
the transformation in the ``neighborhood'' of the inverse square 
potential
amounts to $\beta=-(2-\epsilon)<0$, with $\epsilon \ll 1$,
which implies that $\widetilde{\beta}= 4/\epsilon -2 \gg 1$
and $\alpha = 2/\epsilon$, thus
 converting Eq.~(\ref{eq:radialSchr_powerlaw3}) into
\begin{equation}
\left[
\frac{d^{2}}{dz^{2}} +
\frac{4}{\epsilon^{2}}  \lambda  \,
 \kappa^{-\epsilon}
+ \sigma  \, \frac{4 }{\epsilon^{2}}  z^{4/\epsilon -2 }
- \frac{(\widetilde{l} +
           \widetilde{\nu})^{2} - 1/4}{z^{2}}
\right]
w_{l}(z) = 0
\;  ,
\label{eq:radialSchr_ISP}
\end{equation}
where the potential energy  term
$ - \sigma \, 4
z^{4/\epsilon -2}/\epsilon^{2}$ is seen to behave as an infinite
 hyperspherical  potential well in the limit $\epsilon= 0^{+}$,
for $E<0$ (i.e., $\sigma=-1$).

\bigskip

\begin{center}
{\small\bf ACKNOWLEDGEMENTS}
\end{center}

This research was supported in part by
CONICET and ANPCyT, Argentina
(L.N.E., H.F.,
and C.A.G.C.) and by
the University of San Francisco Faculty Development Fund
(H.E.C.).
H.E.C. acknowledges
the generous hospitality of the University of Houston
while this article was written.


\begin{references}
\bibitem{cam:00_delta}
An extensive list of references for the delta-function potential,
in addition to Refs.~\cite{tho:79,jac:91,alb:88},
can be found in Ref.~\cite{cam:00a}.
\bibitem{cam:00b}
A list of references for this and related
issues can be found in Ref.~\cite{cam:00a}.
\bibitem{cam:00c}
This is very similar to the alternative
procedure carried out in Ref.~\cite{cam:00a} for the
two-dimensional delta-function potential.
\bibitem{cam:00d}
A straightforward derivation can be found in Appendix A
of Ref.~\cite{cam:00a}.
\bibitem{cam:00e}
The generalization of the usual three-dimensional results can be found 
in Appendix C
of Ref.~\cite{cam:00a}.
\end{references}
\end{document}